\documentstyle[12pt]{article}

\catcode`\@=11
\def\marginnote#1{}
\newcount\hour
\newcount\minute
\newtoks\amorpm
\hour=\time\divide\hour by60 \minute=\time{\multiply\hour by60
\global\advance\minute by-\hour}\edef\standardtime{{\ifnum\hour<12
\global\amorpm={am}%
        \else\global\amorpm={pm}\advance\hour by-12 \fi
        \ifnum\hour=0 \hour=12 \fi
        \number\hour:\ifnum\minute<10
0\fi\number\minute\the\amorpm}}
\edef\militarytime{\number\hour:\ifnum\minute<10
0\fi\number\minute}

\def\draftlabel#1{{\@bsphack\if@filesw {\let\thepage\relax
   \xdef\@gtempa{\write\@auxout{\string
      \newlabel{#1}{{\@currentlabel}{\thepage}}}}}\@gtempa
   \if@nobreak \ifvmode\nobreak\fi\fi\fi\@esphack}
        \gdef\@eqnlabel{#1}}
\def\@eqnlabel{}
\def\@vacuum{}
\def\draftmarginnote#1{\marginpar{\raggedright\scriptsize\tt#1}}
\def\draft{\oddsidemargin -.5truein
        \def\@oddfoot{\sl preliminary draft \hfil
        \rm\thepage\hfil\sl\today\quad\militarytime}
        \let\@evenfoot\@oddfoot \overfullrule 3pt
        \let\label=\draftlabel
        \let\marginnote=\draftmarginnote

\def\@eqnnum{(\theequation)\rlap{\kern\marginparsep\tt\@eqnlabel}%
\global\let\@eqnlabel\@vacuum}  }

\def\numberbysection{\@addtoreset{equation}{section}
        \def\theequation{\thesection.\arabic{equation}}}

\def\underline#1{\relax\ifmmode\@@underline#1\else
 $\@@underline{\hbox{#1}}$\relax\fi}

\catcode`@=12 \relax

\numberbysection

\topmargin by -\headsep

\def\rf#1{(\ref{#1})}
\def\lab#1{\label{#1}}

\def\br{\begin{eqnarray}}
\def\er{\end{eqnarray}}
\def\be{\begin{equation}}
\def\ee{\end{equation}}
\def\eq{\!\!\!\! &=& \!\!\!\! }

\def\({\left(}
\def\){\right)}

\newcommand{\bi}[1]{\bibitem{#1}}

\relax

\def\a{\alpha}

\def\d{\delta}
\def\D{\Delta}

\def\l{\lambda}
\def\L{\Lambda}

\def\o{\over}

\def\O{\Omega}

\def\pa{\partial}

\def\ra{\rightarrow}

\def\tp0{\Theta_{+}^{(0)}}
\def\tm0{\Theta_{-}^{(0)}}
\def\ti{\tilde}

\def\vp{\varphi}

\def\nn{\nonumber}

\def\f#1#2#3 {f^{#1#2}_{#3}}

\def\win1{{\sf w_{1+\infty}}}

\def\Win1{{\sf W_{1+\infty}}}

\def\rlx{\relax\leavevmode}
\def\inbar{\vrule height1.5ex width.4pt depth0pt}
\def\IZ{\rlx\hbox{\sf Z\kern-.4em Z}}
\def\IR{\rlx\hbox{\rm I\kern-.18em R}}
\def\IC{\rlx\hbox{\,$\inbar\kern-.3em{\rm C}$}}
\def\IN{\rlx\hbox{\rm I\kern-.18em N}}
\def\IO{\rlx\hbox{\,$\inbar\kern-.3em{\rm O}$}}
\def\IP{\rlx\hbox{\rm I\kern-.18em P}}
\def\IQ{\rlx\hbox{\,$\inbar\kern-.3em{\rm Q}$}}
\def\IF{\rlx\hbox{\rm I\kern-.18em F}}
\def\IG{\rlx\hbox{\,$\inbar\kern-.3em{\rm G}$}}
\def\IH{\rlx\hbox{\rm I\kern-.18em H}}
\def\II{\rlx\hbox{\rm I\kern-.18em I}}
\def\IK{\rlx\hbox{\rm I\kern-.18em K}}
\def\IL{\rlx\hbox{\rm I\kern-.18em L}}
\def\one{\hbox{{1}\kern-.25em\hbox{l}}}
\def\0#1{\relax\ifmmode\mathaccent"7017{#1}%
B        \else\accent23#1\relax\fi}

                \def\JHEP#1#2#3{{\sl JHEP} {\bf#1} (#2) #3}
                \def\PRL#1#2#3{{\sl Phys. Rev. Lett.} {\bf#1} (#2) #3}
                \def\NPB#1#2#3{{\sl Nucl. Phys.} {\bf B#1} (#2) #3}
                
                \def\CMP#1#2#3{{\sl Commun. Math. Phys.} {\bf #1} (#2) #3}
                \def\PRD#1#2#3{{\sl Phys. Rev.} {\bf D#1} (#2) #3}

                \def\PLB#1#2#3{{\sl Phys. Lett.} {\bf #1B} (#2) #3}
                \def\JMP#1#2#3{{\sl J. Math. Phys.} {\bf #1} (#2) #3}
                \def\JNMP#1#2#3{{\sl J. Nonl. Math. Phys.} {\bf #1} (#2) #3}
                \def\PTP#1#2#3{{\sl Prog. Theor. Phys.} {\bf #1} (#2) #3}
                
                \def\AoP#1#2#3{{\sl Annals Phys.} {\bf #1} (#2) #3}
                
                \def\RMP#1#2#3{{\sl Rev. Mod. Phys.} {\bf #1} (#2) #3}

                \def\LMP#1#2#3{{\sl Letters in Math. Phys.} {\bf #1} (#2) #3}
                \def\IJMPA#1#2#3{{\sl Int. J. Mod. Phys.} {\bf A#1} (#2) #3}

                \def\TMP#1#2#3{{\sl Theor. Mat. Phys.} {\bf #1} (#2) #3}
                \def\JPA#1#2#3{{\sl J. Physics} {\bf A#1} (#2) #3}
                
                \def\MPLA#1#2#3{{\sl Mod. Phys. Lett.} {\bf A#1} (#2) #3}

                \def\JPIV#1#2#3{{\sl J. Phys. IV} {\bf #1} (#2) #3}

                \def\a{\alpha}

                \def\d{\delta}
                \def\D{\Delta}

                \def\vp{\varphi}

                \def\/{\frac}

                \def\k{\kappa}
                \def\l{\lambda}
                \def\L{\Lambda}

                \def\o{\omega}
                \def\O{\Omega}
                
                \def\pa{\partial}
                
                \def\qq{\qquad}
                \def\ra{\rightarrow}
                
                \def\vp{\varphi}

                \def\ti{\tilde}
                \def\u{\upsilon}

                \def\({\Big(}
                \def\){\Big)}
                \def\[{\Big[}
                \def\]{\Big]}

                \def\rlx{\relax\leavevmode}
                \def\inbar{\vrule height1.5ex width.4pt depth0pt}
                \def\IZ{\rlx\hbox{\sf Z\kern-.4em Z}}
                \def\IR{\rlx\hbox{\rm I\kern-.18em R}}
                \def\IC{\rlx\hbox{\,$\inbar\kern-.3em{\rm C}$}}
                \def\IN{\rlx\hbox{\rm I\kern-.18em N}}
                \def\IO{\rlx\hbox{\,$\inbar\kern-.3em{\rm O}$}}
                \def\IP{\rlx\hbox{\rm I\kern-.18em P}}
                \def\IQ{\rlx\hbox{\,$\inbar\kern-.3em{\rm Q}$}}
                \def\IF{\rlx\hbox{\rm I\kern-.18em F}}
                \def\IG{\rlx\hbox{\,$\inbar\kern-.3em{\rm G}$}}
                \def\IH{\rlx\hbox{\rm I\kern-.18em H}}
                \def\II{\rlx\hbox{\rm I\kern-.18em I}}
                \def\IK{\rlx\hbox{\rm I\kern-.18em K}}
                \def\IL{\rlx\hbox{\rm I\kern-.18em L}}
                \def\one{\hbox{{1}\kern-.25em\hbox{l}}}
                \def\0#1{\relax\ifmmode\mathaccent"7017{#1}%
                B        \else\accent23#1\relax\fi}
                
\begin{document}           
\begin{titlepage}
\begin{center}
  {\large\bf Generalized sine-Gordon and massive Thirring
  models}\footnote{To appear as a Chapter in {\bf Progress in Soliton Research}
  (Nova Science Publishers, 2004).}
\end{center}

\begin{center}
Harold Blas\\
\vspace{.4 cm}
Departamento de Matem\'atica - ICET\\
Universidade Federal de Mato Grosso (UFMT)\\
 Av. Fernando Correa, s/n, Coxip\'o \\
78060-900, Cuiab\'a - MT - Brazil
\end{center}


\setcounter{page}{1}

\begin{abstract}
We consider the Lagrangian description of the soliton sector of the so-called
affine $\hat{sl}(3)$ Toda model coupled to matter (Dirac) fields (ATM). The
theory is treated as a constrained system in the contexts of the
Faddeev-Jackiw, the symplectic, as well as the master Lagrangian
approaches. We exhibit the master Lagrangian nature of the model from which
generalizations of the sine-Gordon (GSG) or the massive Thirring (GMT) models
are derivable. The GMT model describes $N_{f}=3$ [number of positive roots of
$su(3)$] massive Dirac fermion species with current-current interactions
amongst all the $U(1)$ species currents; on the other hand,  the GSG theory
corresponds to  $N_{b}=2$ [rank of the $su(3)$ Lie algebra] independent Toda
fields (bosons) with a potential given by the sum of three SG cosine terms. The
dual description of the model is further emphasized by providing  some on
shell relationships between bilinears of the GMT spinors and the relevant
expressions of the GSG fields. In this way, in the {\bf first part} of the
chapter, we exhibit the strong/weak coupling phases and the (generalized)
soliton/particle correspondences of the model at the classical level. In the
{\bf second  part} of the chapter we give a full Lie algebraic formulation of
the duality at the level of the equations of motion written in matrix
form. The effective off-critical $\hat{sl}(3)$ ATM action is written in terms of the
Wess-Zumino-Novikov-Witten (WZNW) action plus some kinetic terms for the
spinors  and scalar-spinor interaction terms. Moreover, this theory still presents a remarkable
equivalence between the Noether and topological currents,
describes the soliton sector of the original model and turns out to be the
master Lagrangian describing the GMT and GSG models.
\end{abstract}
\end{titlepage}

\section{Introduction}
\label{sec:int}

Integrable theories in two-dimensions have been an extraordinary laboratory
for the understanding of basic nonperturbative aspects of physical theories
and various aspects, relevant in more realistic $4$-dimensional models, have
been tested \cite{abdalla}. In particular the conformal affine Toda models
coupled to (Dirac) matter fields (CATM) \cite{matter} for the $sl(2)^{(1)}$
and $sl(3)^{(1)}$ cases are discussed in \cite{bla,nucl, annals} and \cite{bueno, jmp,
jhep}, respectively. The interest in such models comes from their
integrability and duality properties \cite{matter,nucl}, which can be used as
toy models to understand some phenomena; such as, a confinement mechanism in
QCD \cite{bla, bueno, prd}. The corresponding off-critical sub-models, such as the $\hat{sl}(2)$ affine Toda model
coupled to matter field (ATM) may  describe some low dimensional condensed matter phenomena, such as self-trapping of electrons into solitons, see e.g. \cite{brazovskii}, tunneling in the integer quantum Hall effect \cite{barci}, and, in particular, polyacteline molecule systems in connection with fermion number fractionization \cite{jackiw}.

The off-critical $\hat{sl}(2)$ ATM model can be obtained at the classical or quantum mechanical
level through some convenient reduction processes starting from CATM
\cite{nucl,annals}. In the $sl(2)$ case, using bosonization techniques, it has
been shown that the classical equivalence between the $U(1)$ vector and
topological currents holds true at the quantum level, and then leads to a bag
model like mechanism for the confinement of the spinor fields inside the
solitons; in addition, it has been shown that the $sl(2)$ ATM theory decouples
into a sine-Gordon model (SG) and a free scalar \cite{bla,witten}. These facts
indicate the existence of a sort of duality in these models involving solitons
and particles. In ref. \cite{annals} through the Faddeev-Jackiw (FJ)\cite{ja}
and ``symplectic quantization'' \cite{symplectic, montani} methods and imposing the
equivalence between the $U(1)$ vector and topological currents as a constraint
 it has been recovered the usual  massive Thirring (MT) and sine-Gordon (SG)
 models equivalence at the classical level; in particular, the mappings between spinor bilinears of the MT theory and exponentials of the SG fields were
                established on shell and the soliton/particle correspondence
                were uncovered \cite{nucl}. Moreover, it has recently been shown that the $\hat{sl}(2)$ ATM model describes a
confinement mechanism and the low-energy spectrum of QCD$_{2}$ (one flavor and
$N$ colors) \cite{prd}.

One of the difficulties with generalizations of complex affine Toda field
theories, beyond $su(2)$ and its associated SG model, has to do with
unitarity. In this chapter we describe the many field generalizations of SG/MT models
based on soliton/particle duality and unitarity. Beyond the well known $sl(2)$
case the related $sl(n)^{(1)}$ CATM model does not possess a local real Lagrangian,
therefore it is defined an off-critical sub-model Lagrangian with well behaved
classical solutions making use of the results of \cite{bueno}. In \cite{bueno}
the authors studied the $sl(3)^{(1)}$ CATM soliton solutions and some of their
properties up to general $2$-soliton. Using the FJ and symplectic methods it
has been show the parent Lagrangian \cite{hje} nature of the $sl(3)$ ATM model
from which the generalized sine-Gordon (GSG) or the massive Thirring (GMT)
models are derivable \cite{jmp, jhep}. It is then shown that there are (at least classically) two
equivalent descriptions of the model, by means of either the Dirac or the Toda
type fields. It is also clear the duality exchange of the coupling regimes $g
\rightarrow 1/g$ in each
$\hat{sl}(2)$ ATM sub-model, in this way  generalizing the soliton/particle correspondences which is uncovered by providing explicit relationships
between the GSG and GMT fields. The ordinary MT/SG duality turns out to be related to each
$SU(2)$ sub-group. The $\hat{sl}(n)$ affine Lie
algebra generalizations are outlined following \cite{jhep}. In this way we describe a precise field content of both sectors; namely, the correct GMT/GSG duality, first undertaken in \cite{halpern}.

Recently, the three fermion species GMT model has been bosonized and the
 $\hat{sl}(3)$ ATM model emerged
 in the semi-classical limit of an intermediate Lagrangian \cite{bosoni}. The generalized Mandelstam soliton operators were
constructed and the fermion-boson mapping was established through a set of
generalized bosonization rules in a quotient positive definite Hilbert space
of states.

The chapter is organized as follows. In section 2 we define the $\hat{sl}(3)$ ATM
model . Section 3 deals with the model in the FJ framework \cite{ja}, the outcome is the
GMT model. In section 4, we attack the same problem from the point of view of
symplectic quantization \cite{symplectic,montani} giving the Poisson brackets
of the GMT and GSG models. Section 5 deals with the soliton/particle and
strong/weak coupling correspondences. Section 6 provides a full Lie algebraic
construction of the relevant results. Section 7 outlines the relevant steps towards the generalization to $\hat{sl}(n)$ ATM.\, In the appendix \ref{app:catm} we present the construction of $sl(3)^{(1)}$ CATM model.

\section{The model}
\label{sec:ATM}

In affine Toda type theories the question of whether all mathematical solutions are physically acceptable deserves a careful analysis, specially if any consistent quantization of the models is discussed. The requirement of real energy density leads to a certain reality conditions on the solutions of the model. In general, a few soliton solutions survive the reality constraint, if in addition one also demands positivity. These kind of issues are discussed in refs. \cite{hermitian}. Here we follow the prescription to restrict the model to a subspace of classical solutions which satisfy the physical principles of reality of energy density and soliton/particle correspondence.

In CATM models associated to the principal gradation of an affine Lie algebra,
it is a matter of fact that a $1$-soliton real solution for the Toda field
corresponds to each pair of Dirac fields $\psi^{i}$\, and
$\widetilde{\psi}^{i}$. This fact allows us to make the identifications
$\widetilde{\psi}^{i}\, \sim \,(\psi^{i})^{*}$($\star$ means complex
conjugation), and take real Toda fields. In the case of $sl(2)^{(1)}$ CATM theory, this procedure does not spoil the particle-soliton correspondence \cite{bla,nucl}.

We consider the $sl(3)^{(1)}$ CATM theory (see Appendix \ref{app:catm}) with the conformal symmetry gauge fixed \footnote{The auxiliary fields $\widetilde{\nu}$ and $\eta$ of the CATM theories are associated to the topological character of the soliton masses and to the conformal symmetry, respectively. The classical and quantum reductions CATM $\rightarrow$ ATM can be treated as in \cite{annals} and \cite{nucl}, respectively.} by setting $\eta\,=\,0$ and the reality conditions
\br
\lab{real1}
\widetilde{\psi}^{j}&=&-{(\psi^{j})}^{*},\,\, (j=1,2,3);\,\,\,\,\vp_{a}^{*}= \vp_{a},\,\, (a=1,2),\\
\nonumber \mbox{or}
\\
\widetilde{\psi}^{j}&=&{(\psi^{j})}^{*},\,\,
j=1,2;\,\,\,\,\,\,\,\widetilde{\psi}^{3}=-{(\psi^{3})}^{*},\nonumber
\\\, \,\, \vp_{1,\,2}&\rightarrow& \vp_{1,\,2}- \pi \,\,\,\,\,\, (\mbox{the new}\,\vp_{a}\mbox{'s being real fields}).\lab{real2}
\er

The condition \rf{real2} must be supplied with $x^{\mu}\rightarrow -x^{\mu}$.
Moreover, for consistency of the equations of motion \rf{eqnm4}-\rf{eqnm15} under the reality conditions \rf{real1}-\rf{real2}, from eqs. \rf{eqnm5}, \rf{eqnm7}, \rf{eqnm8}, \rf{eqnm10}, \rf{eqnm12} and \rf{eqnm15}, we get the relationships
\br
\widetilde{\psi }_{L}^{j}\psi _{R}^{3}-\widetilde{\psi }_{R}^{j}\psi _{L}^{3}e^{-3i\varphi_{j}}\,=\,0,\,\,j=1,2;\,\,\,\, \psi _{L}^{1}\psi_{R}^{2}e^{-3i\varphi_{1}}-\psi _{L}^{2}\psi_{R}^{1}e^{-3i\varphi_{2}}\,=\,0. \lab{condi}
\er

Then, the above reality conditions and constraints allow us to define a suitable physical Lagrangian. The equations \rf{eqnm1}, \rf{eqnm4}-\rf{eqnm15}, supplied with \rf{real1} [or \rf{real2}] and \rf{condi}, follow from the Lagrangian
\br
\lab{atm1}
\frac{1}{k}{\cal L} = \sum_{j=1}^{3} \[ \frac{1}{24}\partial_{\mu }\phi_j \partial ^{\mu }\phi_j + i\overline{\psi}^j \gamma ^{\mu}\partial _{\mu }\psi^j - m^{j}_{\psi }\overline{\psi}^j e^{i \phi_j  \gamma_{5}}\psi^j\]
\er
where ${\bar{\psi}}^{j} \equiv {({\psi}^{j})}^{\dagger} \,\gamma_0\,$,\, $\phi_{1}\equiv 2\vp_{1}-\vp_{2}$,\, $\phi_{2}\equiv 2\vp_{2}-\vp_{1}$,\, $\phi_{3}\equiv \phi_{1}+\phi_{2}$,\, $m_{\psi}^3=m_{\psi}^1+m_{\psi}^2$, \,$k$ is an overall coupling constant and the $\vp_{j}$ are real fields.

The eq. \rf{atm1} defines the  $sl(3)$ {\sl affine Toda theory coupled to matter fields} (ATM). Notice that the space of solutions of $sl(3)^{(1)}$ CATM model satisfying the conditions \rf{real1}-\rf{condi} must be solutions of the $sl(3)$ ATM theory \rf{atm1}. Indeed, it is easy to verify that the three species of one-soliton solutions [$S\equiv${\sl 1-soliton}($\bar{S}\equiv${\sl 1-antisoliton})] \cite{bueno}: $\{\(\vp_{1},\, \psi^{1}\)_{S/\bar{S}},\, \vp_{2}=0,\,\psi^{2}=0,\,\psi^{3}=0\}$, $\{\(\vp_{2},\, \psi^{2}\)_{S/\bar{S}},\, \vp_{1}=0,\,\psi^{1}=0,\,\psi^{3}=0\}$ and  $\{\(\vp_{1}+\vp_{2},\,\psi^{3}\)_{S/\bar{S}},\, \vp_{1}=\vp_{2},\,\psi^{1}=0,\,\psi^{2}=0\}$\, satisfy the equations of motion; i.e., each positive root of $sl(3)$ reproduces the $sl(2)$ ATM case \cite{bla,nucl}. Moreover, these solutions satisfy the above reality conditions and constraints \rf{real1}-\rf{condi} (with \rf{real1} and \rf{real2} for $S$ and $\bar{S}$, respectively), and the equivalence between the $U(1)$ vector and topological currents \rf{equivalence}. Then, the soliton/particle correspondences survive the above reduction processes performed to define the $sl(3)$ ATM theory.

The class of $2$-soliton solutions of $sl(3)^{(1)}$ CATM \cite{bueno} behave as  follows:\, i) they are given by 6 species associated to the pair $(\a_{i},\a_{j}),\, i\le j;\,\, i,j=1,2,3$; where the $\a$'s are the positive roots of $sl(3)$ Lie algebra. Each species $(\a_{i},\a_{i})$ solves the $sl(2)$ CATM submodel\footnote{$sl(2)$ ATM $2-$solitons satisfy an analogous eq. to \rf{equivalence}; for $\vp$ real and $\widetilde{\psi}=\pm (\psi)^{*}$ (constraints \rf{real1}-\rf{real2}; \rf{condi} is trivialy satisfied since $\widetilde{\psi}_{j}=\psi_{j}=0$ for $j\ne i$) one has, soliton-soliton $SS$, $SS$ bounds and no $S\bar{S}$ bounds \cite{bla}.};
ii) satisfy the $U(1)$ vector and topological currents equivalence \rf{equivalence}.

\section{The generalized massive Thirring model (GMT)}
\label{sec:FJ}

Let us consider the following Lagrangian
\be
\lab{lagrangian1}
       \frac{1}{k} {\cal L} = \sum_{j=1}^{3}\Big[\frac{1}{24}\partial_{\mu }\phi_j \partial ^{\mu }\phi_j
        + i\overline{\psi}^j \gamma ^{\mu}\partial _{\mu }\psi^j - m^{j}_{\psi }\overline{\psi}^j e^{i \phi_j  \gamma_{5}}\psi^j
+ \lambda^{j}_{\mu} ( m^{j } \overline{\psi}^{j}\gamma^{\mu}\psi^{j} - \epsilon^{\mu\nu}\partial_{\nu} (q_j \phi_j))\Big],
\ee
where the ATM Lagrangian \rf{atm1} is supplied with the constraints,
$ ( m^{l} \overline{\psi}^{l}\gamma^{\mu}\psi^{l}+\frac{m^{3}}{2} \overline{\psi}^{3}\gamma^{\mu}\psi^{3} - \epsilon^{\mu\nu}\partial_{\nu} \phi_{l})$, ($l=1,2$), with the help of the Lagrange multipliers
$\lambda^{j}_{\mu}$\, ($\lambda^{3}_{\mu} \equiv
\frac{\lambda^{1}_{\mu}+\lambda^{2}_{\mu}}{2} $,\, $q_{1}\equiv q_{2} \equiv
1$,\, $q_{3}\equiv 0$). Their total sum bears an intriguing resemblance to the
$U(1)$ vector and topological currents equivalence \rf{equivalence}; however,
the $m^{j}$'s here are some arbitrary parameters. The same procedure has been
used, for example, to incorporate the left-moving condition in the study of
chiral bosons in two dimensions \cite{siegel}. The Lagrangian
\rf{lagrangian1} still possesses the left-right local symmetries
\rf{leri1}-\rf{leri2} of $sl(3)$ ATM \rf{atm1} provided a convenient
transformation for the field $\l_{j}^{\mu}$ is assumed\footnote{In \cite{jmp}
  this feature has not been pointed out.}. In order to apply the Faddeev-Jackiw (FJ) method we should write \rf{lagrangian1} in the first order form in time derivative, so let us define the conjugated momenta
\br
\lab{moments}
\pi _{1}&\equiv& \pi_{\phi_{1}}\,=\,\frac{1}{12} ( 2 \dot{\phi}_1 + \dot{\phi}_2) + \lambda^{1}_{1},
        \qquad
\pi _{2} \equiv \pi_{\phi_{2}}\,=\,\frac{1}{12} ( 2 \dot{\phi}_2 + \dot{\phi}_1) + \lambda^{2}_{1} ,   \nonumber
        \\
\pi _{\lambda^{1}_{\mu}}&=& 0,
        \qquad
\pi _{\lambda^{2}_{\mu}} = 0,
        \qquad
\pi^{j}_{R}\equiv \pi^{j}_{\psi_{R}} =-i{\ti \psi}^{j}_{R},
        \qquad
\pi^{j}_{L} \equiv \pi^{j}_{\psi_{L}} =-i{\ti \psi}^{j}_{L}.
\er

We are assuming that Dirac fields are anticommuting Grasmannian variables and their momenta variables defined through {\bf left} derivatives.
Then, as usual, the Hamiltonian is defined by (sum over repeated indices is assumed)
\be
\lab{hamiltonian}
        {\cal H}_{c} = \pi_{1} \dot{\phi}_{1} + \pi_{2} \dot{\phi}_{2}
+ \dot{\psi}^{j}_{R} \pi^{j}_{R}  +  \dot{\psi}^{j}_{L} \pi^{j}_{L}     - \cal{L}.
\ee

Explicitly the Hamiltonian density becomes
\begin{eqnarray}
\lab{hamiltonian1}
{\mathcal{H}}_{c} &=& 2 ( \pi_{j})^{2} + 4 ( \lambda^{1}_{1})^{2}
+ 4\left ( \lambda^{2}_{1}\right)^{2} - \lambda^{1}_{1}
{\mathcal{J}}_{1} - \lambda^{2}_{1} {\mathcal{J}}_{2} - 4 \left( \lambda^{1}_{1} \lambda^{2}_{1} \right)  \nonumber
        \\
&&+\frac{1}{24}\left( \phi _{j,x}\right) ^{2}-\pi _{R}^{j}\psi_{R,x}^{j} +
        \pi _{L}^{j}\psi _{L,x}^{j}+im_{\psi }^{j}(e^{-\phi _{j}}
        \tilde{\psi}_{R}^{j}\psi _{L}^{j}-e^{\phi _{j}}\tilde{\psi}_{L}^{j}\psi _{R}^{j})
\nonumber
\\
&&+\lambda^{1}_{0}[J_{1}^{0}-\phi _{1,x}]
+\lambda^{2}_{0}[J_{2}^{0}-\phi _{2,x}],
\end{eqnarray}
where $\pi _{3} \equiv \pi _{1}-\pi _{2}$,\, ${\mathcal{J}}_{1} \equiv J_{1}^{1}+ 4\left( 2\pi_{1}-\pi _{2}\right)$,\,\,
        ${\mathcal{J}}_{2} \equiv J_{2}^{1}
        + 4\left( 2\pi _{2}-\pi _{1}\right)$\,
 and
\br
J_{1}^{\mu}\,=\,m^{1} j^{\mu}_{l}+\frac{m^{3}}{2} j^{\mu}_{3};\,\,\,\,J_{2}^{\mu}\,=\,m^{2} j^{\mu}_{2}+\frac{m^{3}}{2} j^{\mu}_{3};\,\,\,\,\,\,j^{\mu}_{l}\,\equiv\, \bar{\psi}^{l}\gamma^{\mu}\psi^{l},\,\,\,\,l=1,2,3.
\er

Let us observe that each $U(1)$ Noether current of the $sl(3)$ ATM theory defined in \rf{atm1} is conserved separately; i.e., $\pa_{\mu}j^{\mu}_{l}=0,\,\,\,\, l=1,2,3 $.

Next, the same Legendre transform \rf{hamiltonian} is used to write the first order Lagrangian
\br
\lab{lagran}
        {\cal L}& = \pi_{1}\dot{\phi}_{1}  + \pi_{2}\dot{\phi}_{2}
        +  \dot{\psi}^{j}_{R} \pi^{j}_{R}  + \dot{\psi}^{j}_{L} \pi^{j}_{L}
        - {\cal H}_{c}.
\er

Our starting point for the FJ analysis will be this first order Lagrangian. Then the Euler-Lagrange equations for the Lagrange multipliers allow one to solve two of them
\br
\lambda^{1}_{1}= \frac{2 {\mathcal{J}}_{1} +  {\mathcal{J}}_{2} }
                {12 }, \qq
\lambda^{2}_{1}=  \frac{2  {\mathcal{J}}_{2} +  {\mathcal{J}}_{1} }
                {12 }
\er
and the remaining equations lead to two constraints
\be
\lab{cons}
        \O_{1} \equiv J_{1}^{0}-\phi _{1,x} = 0,
        \qquad
        \O_{2} \equiv J_{2}^{0}-\phi _{2,x} = 0.
\ee

The Lagrange multipliers $\lambda^{1}_{1}$ and $\lambda^{2}_{1}$ must be
replaced back in \rf{lagran} and the constraints \rf{cons} solved. Firstly,
let us replace the $\lambda^{1}_{1}$ and $\lambda^{2}_{1}$ multipliers into
${\cal H}_{c}$, then one gets
\br
\lab{hamiltonian2}
{\cal H}_{c}^{\prime} &=& 2(\pi_{j})^{2} - \frac{1}{12}
        \{ ( {\cal J}_1)^2 + ( {\cal J}_2)^2
        +({\cal J}_1 {\cal J}_2)\} + \frac{1}{24} ( \phi_{j,x})^2
        \nonumber
        \\
        &+& i \tilde{\psi}^{j}_{R} {\psi}^{j}_{R,x}
         -  i \tilde{\psi}^{j}_{L} {\psi}^{j}_{L,x}
        + im^{j}_{\psi} ( e^{-i\phi_{j}} \tilde{\psi}^{j}_{R}\psi^{j}_{L}
                        - e^{ i\phi_{j}} \tilde{\psi}^{j}_{L}\psi^{j}_{R}).
\er

The new Lagrangian becomes
\be
\lab{lagran1}
        {\cal L}^{\prime} = \pi_{1}\dot{\phi}_1 + \pi_{2}\dot{\phi}_2
        + \dot{\psi}^{j}_{R} \pi^{j}_{R}
        + \dot{\psi}^{j}_{L} \pi^{j}_{L} - {\cal H}_{c}^{\prime}.
\ee

We implement the constraints \rf{cons} by replacing in \rf{lagran1} the fields
$\phi_{1},\,\phi_{2}$ in terms of the space integral of the current components $J^{0}_{1},\,J^{0}_{2}$. Then we get the Lagrangian
\begin{eqnarray}
\lab{lagran2}
{\cal L}^{\prime\prime} &=& \pi_{1} \pa_{t} \int^{x} J^{0}_{1}
                                  + \pi_{2} \pa_{t} \int^{x} J^{0}_{2}
                                  + \dot{\psi}^{j}_{R} \pi^{j}_{R}
                                + \dot{\psi}^{j}_{L} \pi^{j}_{L}
         - i {\tilde \psi}^{j}_{R} \psi_{R,x}^{j} + i{\tilde \psi}^{j}_{L} {\psi}_{L,x}^{j}
         \\                                                                                             \nonumber
         &-& i m^{j}_{\psi} \(  e^{-i \int^{x} J^{0}_{j}} {\tilde \psi}^{j}_{R} \psi^{j}_{L}
                            - e^{ i \int^{x} J^{0}_{j}} {\tilde \psi}^{j}_{L} \psi^{j}_{R}  \)
         - \frac{1}{12}  \((J_{1})^2 + (J_{2})^2
         + J_{1}.J_{2} \)
         \\
         \nonumber
         &+& \pi_1 J^{1}_{1} +  \pi_2 J^{1}_{2},
\end{eqnarray}
where $J^{0}_{3} \equiv J^{0}_{1}+J^{0}_{2}$. Observe that the terms containing the $\pi_{a}$'s in eq. \rf{lagran2} cancel to each other if one uses the current conservation laws. Notice the appearances of various types of current-current interactions. The following Darboux transformation
\br
\lab{darboux}
\psi^{j}_{R} \rightarrow e^{-\frac{i}{2} \int^{x} J^{0}_{j}} \psi^{j}_{R},
        \qquad
        \psi^{j}_{L} \rightarrow e^{    \frac{i}{2} \int^{x} J^{0}_{j}} \psi^{j}_{L},\,\,\,\,\,\,j=1,2,3,
\er
is used to diagonalize the canonical one-form. Then, the kinetic terms will give additional current-current interactions, $-\frac{1}{2}[J_{1}.(j_{1}+j_{3})+ J_{2}.(j_{2}+j_{3})]$.  We are, thus, after defining $k\equiv 1/g$, and rescaling the fields $\psi^{j} \rightarrow 1/\sqrt{k}\, \psi^{j}$, left with the Lagrangian
\begin{equation}
\label{thirring1}
{\cal L}[\psi,\overline{\psi}]= \sum_{j=1}^{3}\{i\overline{\psi}^{j}\gamma^{\mu}\pa_{\mu}\psi^{j}
        + m^{j}_{\psi}\,\,{\overline{\psi}}^{j}\psi^{j}\}\,
        \,- \sum_{{\begin{array}{c} k,l=1\\k\leq l\end{array}}}^{3}\[\bar{a}_{kl}j_{k}.j_{l}\],
\end{equation}
where $\bar{a}_{kl}=g\, a_{kl}$, with
$a_{33}=\frac{1}{2}(\frac{(m^{3})^2}{8}+m^{3})$;
$a_{12}=\frac{1}{12}m^{1}m^{2}$;
$a_{ii}=\frac{1}{2}(\frac{(m^{i})^2}{6}+m^{i})$;
$a_{i3}=\frac{1}{2}(\frac{m^{i} m^{3}}{4}+m^{i}+ \frac{m^{3}}{2})$,
$i=1,2$. This defines the {\sl generalized massive Thirring model} (GMT)
\cite{jmp, jhep}. The
canonical pairs are  $(-i{\tilde \psi}_{R}^{j}, \psi_{R}^{j})$ and $(-i{\tilde
  \psi}_{L}^{j}, \psi_{L}^{j})$. Let us comment on the integrability and the Grassmannian nature of
the $\psi$ fields. The usual MT model (with Grassmannian fields) classical integrability
has been established by means of its  zero-curvature formulation (see
e.g. \cite{izergin}). In our case it would be interesting to establish the GMT
integrability using similar approach.

From the Lie algebraic constructions of \cite{jhep, bosoni} one has all
                $\bar{a}_{jk} > 0,\,\,j,k=1,2,3;\, \mbox{but}\,\,\bar{a}_{12} < 0$. Taking into account the {\sl signs} of the $\bar{a}_{jk}$'s in the model (\ref{thirring1}) one can infer that the fermions of the same species will experience an attractive force. The pair of fermions of species  $1$ and $3$, as well as $2$ and $3$ also experience attractive forces,  whereas the pair of fermions $1$ and $2$ suffer a repulsive force \cite{rajaraman}. These features can also be deduced from the behavior of the time delays due to soliton-soliton interactions in the associated $\hat{su}(3)$ ATM model studied in ref. \cite{bueno}.

\section{The symplectic formalism and the ATM model}

\subsection{The (constrained) symplectic formalism}

\label{sec:symplectic1}

We give a brief overview of the basic notations of symplectic approach \footnote{These are given for point mechanics, the extension to field theory is self evident.}. The geometric structure is defined by the closed (pre)symplectic two-form
\br
f^{(0)}\,=\,\frac{1}{2}f^{(0)}_{ij}(\xi^{(0)}) d{\xi^{(0)}}^{i}\wedge d{\xi^{(0)}}^{j}
\er
where
\br
\lab{form}
f^{(0)}_{ij}(\xi^{(0)})\eq\frac{\pa}{\pa{{\xi}^{(0)}}^{i}}{\bf a}^{(0)}_{j}(\xi^{(0)})-\frac{\pa}{\pa{{\xi}^{(0)}}^{j}}{\bf a}^{(0)}_{i}(\xi^{(0)})
\er
with ${\bf a}^{(0)}(\xi^{(0)})={\bf a}^{(0)}_{j}(\xi^{(0)}) d{\xi^{(0)}}^{j}$ being the canonical one-form defined from the original first order Lagrangian
\br
L^{(0)}dt\,=\, {\bf a}^{(0)}(\xi^{(0)})-V^{(0)}(\xi^{(0)})dt.
\er

The superscript $(0)$ refers to the original Lagrangian, and is indicative of the iterative nature of the computations. The constraints are imposed through Lagrange multipliers which are velocities, and in such case one has to extend the configuration space \cite{symplectic,montani}. The corresponding Lagrangian gets modified and consequently the superscript also changes. The algorithm terminates once the symplectic matrix turns out to be non-singular.

\subsection{The generalized massive Thirring model (GMT)}

\label{sec:symplectic2}

Next, we will consider our model in the framework of the symplectic formalism. Let ${\cal L}^{\prime}$, eq. \rf{lagran1}, be the zeroth-iterated Lagrangian ${\cal L}^{(0)}$. Then the first iterated lagrangian will be
\be
\lab{itera}
        {\cal L}^{(1)} = \pi_{1} {\dot \phi}_{1} + \pi_{2} {\dot \phi}_{2}
        + \dot{\psi}^{j}_{R} \pi^{j}_{R} + \dot{\psi}^{j}_{L} \pi^{j}_{L}
        + \dot{\eta}^{1}\O_{1} + \dot{\eta}^{2}\O_{2} - {\cal V}^{(1)},
\ee
where the once-iterated symplectic potential is defined by
\be
{\cal V}^{(1)}={\cal H}_{c}^{\prime}|_{\O_{1}=\O_{2}=0},
\ee
and the stability conditions of the symplectic constraints, $\O_{1}$ and $\O_2$, under
        time evolution have been implemented by making $\l^{1}_{0} \rightarrow \dot{\eta}^{1}$ and
        $\l^{2}_{0} \rightarrow \dot{\eta}^{2}$. Consider the once-iterated set of symplectic
        variables in the following order
\be
{\bf \xi^{(1)}} = (\eta^{1}, \eta^{2}, \phi_1, \phi_2, \psi^{1}_{R}, \psi^{1}_{L}, \psi^{2}_{R},
        \psi^{2}_{L}, \psi^{3}_{R}, \psi^{3}_{L}, \pi_{1}, \pi_2, \pi^{1}_{R}, \pi^{1}_{L},
        \pi^{2}_{R}, \pi^{2}_{L}, \pi^{3}_{R}, \pi^{3}_{L} ),
\ee
and the components of the canonical one-form
\be
{\bf a^{(1)}} = (\O_{1}, \O_2, \pi_{1}, \pi_2, -\pi^{1}_{R}, -\pi^{1}_{L}, -\pi^{2}_{R},
        -\pi^{2}_{L}, -\pi^{3}_{R}, -\pi^{3}_{L}, 0, 0, 0, 0, 0, 0, 0, 0).
\ee

These result in the singular symplectic two-form $18\mbox{x}18$ matrix
\br
f^{(1)}_{AB}(x,y)&=& \left(\begin{array}{cc}
        a_{11} &  a_{12}                        \\
        a_{21} &  a_{22}
        \end{array}\right)\d(x-y),
\er
where the $9\mbox{x}9$ matrices are
\br
a_{11} = \left(\begin{array}{cccccccccc}
        0 & 0 &  \pa_{x} & 0 & i m^1 {\tilde \psi}^{1}_{R}
                                      & i m^1 {\tilde \psi}^{1}_{L}
        & 0 & 0 & \frac{im^{3}}{2} {\tilde \psi}^{3}_{R}
\\
        0 & 0 & 0 & \pa_{x} & 0 & 0 & i m^2 {\tilde \psi}^{2}_{R}  &
        i m^2 {\tilde \psi}^{2}_{L}  & \frac{im^{3}}{2} {\tilde \psi}^{3}_{R}
        \\
         \pa_x &  0 & 0 & 0 & 0 & 0 & 0 & 0 & 0
        \\
        0 &  \pa_x & 0 &  0 & 0 & 0 & 0 & 0 & 0
        \\
        i m^1 {\tilde \psi}^{1}_{R} & 0 & 0 & 0 & 0 & 0 & 0 & 0 & 0
        \\
        i m^1 {\tilde \psi}^{1}_{L} & 0 &  0 & 0 & 0 & 0 & 0 & 0 & 0
        \\
        0 & i m^2 {\tilde \psi}^{2}_{R} & 0 & 0 & 0 & 0 & 0 & 0 & 0
        \\
        0 & i m^2 {\tilde \psi}^{2}_{L} & 0 & 0 & 0 & 0 & 0 & 0 & 0
        \\
        \frac{im^{3}}{2} {\tilde \psi}^{3}_{R} &
        \frac{im^{3}}{2} {\tilde \psi}^{3}_{R} & 0 & 0 & 0 & 0 & 0 & 0 & 0
        \end{array}\right), \nonumber
\er

\br
a_{12} =  \left(\begin{array}{cccccccccc}
         \frac{im^{3}}{2} {\tilde \psi}^{3}_{L} & 0 & 0
         & m^1 {\psi}^{1}_{R}

& m^1 {\psi}^{1}_{L}
         & 0 & 0 & \frac{m^{3}}{2} {\psi}^{3}_{R} &
                   \frac{m^{3}}{2} {\psi}^{3}_{L}
        \\
        \frac{im^{3}}{2} {\tilde \psi}^{3}_{L} & 0 & 0 & 0 & 0 &
        m^2 { \psi}^{2}_{R} & m^2 {\psi}^{2}_{L} &
        \frac{m^{3}}{2} { \psi}^{3}_{R} &
        \frac{m^{3}}{2} { \psi}^{3}_{L}
        \\
        0 & -1 & 0 & 0 & 0 & 0 & 0 & 0 & 0
        \\
        \vdots & 0 & \ddots & \vdots & \vdots & \vdots & \vdots & \vdots & \vdots\\
0 & 0 & 0 & 0 & 0 & 0 & 0 & -1 & 0
        \end{array}\right), \nonumber
\er
\br
a_{21}&=& \left(\begin{array}{ccccccccc}
         \frac{im^{3}}{2} {\tilde \psi}^{3}_{L} &
         \frac{im^{3}}{2} {\tilde \psi}^{3}_{L} & 0 & 0 & 0 & 0 & 0 & 0 & 0
        \\
        0 & 0 & 1 & 0 & 0 & 0 & 0 & 0 & 0
        \\
        0 & 0 & 0 & 1 & 0 & 0 & 0 & 0 & 0
        \\
        m^1 {\psi}^{1}_{R} & 0 & 0 & 0 & -1 & 0 & 0 & 0 & 0
        \\
        m^1 {\psi}^{1}_{L} & 0 & 0 & 0 & 0 & -1 & 0  & 0 & 0
        \\
        0 & m^2 {\psi}^{2}_{R} & 0 & 0 & 0 & 0 & -1  & 0 & 0
        \\
        0 & m^2 {\psi}^{2}_{L} & 0 & 0 & 0 & 0 & 0 & -1 & 0
        \\
        \frac{m^{3}}{2} {\psi}^{3}_{R} &
        \frac{m^{3}}{2} {\psi}^{3}_{R} & 0 & 0 & 0 & 0 & 0 & 0 & -1
        \\
        \frac{m^{3}}{2} {\psi}^{3}_{L} &
        \frac{m^{3}}{2} {\psi}^{3}_{L} & 0 & 0 & 0 & 0 & 0 & 0 & 0
        \end{array}\right),\nonumber
\\
a_{22}&=& \left(\begin{array}{ccccc}
        0 & 0 & \cdots & 0 & -1\\
        0 & 0 & 0 & 0 & 0\\
        \multicolumn{5}{c}\dotfill
        \\
        0 & 0 & 0 & 0 & 0
        \\
        -1 & 0 & \cdots & 0 & 0
        \end{array}\right). \nonumber
\er

This matrix has the zero modes
\br
        {\bf {{v}}^{(1)}}^{T}(x) &=& \( \frac{-u}{m^1}, \; \frac{-\u}{m^2}, \;
        0, \; 0,  \; u\psi^1_{R}, \; u\psi^1_{L}, \; \u \psi^2_{R}, \; \u \psi^2_{L}, \;
        \frac{m^3}{2} \(\frac{u}{m^1} + \frac{\u}{m^2}\) \psi^3_{R},
        \nonumber
        \\
        && \frac{m^3}{2} \(\frac{u}{m^1} + \frac{\u}{m^2}\) \psi^3_{L}, \;
        -\frac{u^{\prime}}{m^1}, \; -\frac{\u^{\prime}}{m^2}, \;
        i u{\tilde \psi}^1_{R}, \; i u{\tilde \psi}^1_{L}, \;
        i\u{\tilde \psi}^2_{R}, \; i\u{\tilde \psi}^2_{L}, \;
        \nonumber
        \\
        && i\frac{m^3}{2} (\frac{u}{m^1} + \frac{\u}{m^2}) {\tilde \psi}^3_{R}, \;
        \frac{im^3}{2} (\frac{u}{m^1} + \frac{\u}{m^2}) {\tilde \psi}^3_{L} \),
\er
where $u$ and $\u$ are arbitrary functions. The zero-mode condition gives
\be
        \int {dx {\bf{{v}}^{(1)}}^T(x)\frac{\d}{\d \xi^{(1)}(x)}\int{dy{\cal V}^{(1)}}}\equiv 0.
\ee

Thus, the gradient of the symplectic potential happens to be orthogonal to the zero-mode
${\bf {v}}^{(1)}$. Since the equations of motion are automatically validated no symplectic
constraints appear. This happens due to the presence of the symmetries of the action
\begin{eqnarray}
\d \xi^{(1)}_{A}\,=\, {\bf v^{(1)}_{A}}(x); \,\,\,\,A=1,2,...18.
\end{eqnarray}

So, in order to deform the symplectic matrix into an invertible one, we have to add some gauge fixing terms to the symplectic potential. One can choose any consistent set of gauge fixing
conditions \cite{montani}. In our case we have two symmetry generators associated to the
parameters $u$ and $v$, so there must be two gauge conditions. Let us choose
\be
\lab{gauge1}
        \O_{3}\equiv \phi_{1} = 0  \qq  \O_{4}\equiv \phi_{2} = 0.
\ee

These conditions gauge away the fields $\phi_{1}$ and $\phi_{2}$, so only the remaining field variables
will describe the dynamics of the system. Other gauge conditions, which eventually gauge away the spinor fields
$\psi^{i}$ will be considered in the next subsection.

Implementing the consistency conditions by means of Lagrange multipliers $\eta^{3}$ and $\eta^{4}$ we get the twice-iterated Lagrangian
\be
        {\cal L}^{(2)} = \pi_{1}\dot{\phi}_{1} + \pi_{2}\dot{\phi}_{2} + \dot{\psi}_{R}\pi^{j}_{R}
        + \dot{\psi}_{L}\pi^{j}_{L}  + \dot{\eta}^{1}\O_{1} + \dot{\eta}^{2}\O_{2}
        + \dot{\eta}^{3}\O_{3} + \dot{\eta}^{4}\O_{4} - {\cal V}^{(2)},
\ee
where
\br
        {\cal V}^{(2)}\;&=&\;{\cal V}^{(1)}|_{\O_{3} = \O_{4} = 0}.
        \nonumber
\er

Assuming now that the new set of symplectic variables is given in the following order
\be
{\bf \xi^{(2)}} = (\eta^{1}, \eta^{2},\eta^{3}, \eta^{4}, \phi_1, \phi_2, \psi^{1}_{R}, \psi^{1}_{L},
        \psi^{2}_{R}, \psi^{2}_{L}, \psi^{3}_{R}, \psi^{3}_{L}, \pi_{1}, \pi_2, \pi^{1}_{R},
        \pi^{1}_{L},
\pi^{2}_{R}, \pi^{2}_{L}, \pi^{3}_{R}, \pi^{3}_{L} ),
\ee
and the non vanishing components of the canonical one-form
\be
{\bf a^{(2)}} = (\O_1, \O_2, \O_3, \O_4, \pi_{1}, \pi_2, -\pi^{1}_{R}, -\pi^{1}_{L}, -\pi^{2}_{R},
        -\pi^{2}_{L}, -\pi^{3}_{R}, -\pi^{3}_{L}, 0, 0, 0, 0, 0, 0, 0, 0),
\ee
one obtains the singular twice-iterated symplectic $20$x$20$ matrix
\br
f^{(2)}_{AB}(x,y) &=& \left(\begin{array}{cc}
        a_{11} &  a_{12}                        \\
        a_{21} &  a_{22}
        \end{array} \right) \d(x-y),
\er
where the $10$x$10$ matrices are
\br
a_{11} = \left(\begin{array}{cccccccccc}
        0 & 0 &0 & 0 &  \pa_{x} & 0 & i m^1 {\tilde \psi}^{1}_{R}
        & i m^1 {\tilde \psi}^{1}_{L} & 0 & 0
        \\
        0 & 0 & 0 & 0 & 0 & \pa_{x} & 0 & 0 & i m^2 {\tilde \psi}^{2}_{R}  &
        i m^2 {\tilde \psi}^{2}_{L}
        \\
        0 & 0 & 0 & 0 & -1 & 0 & 0 & 0 & 0 & 0
        \\
        0 & 0 & 0 & 0 & 0 & -1 & 0 & 0 & 0 & 0
        \\
         \pa_x & 0 & 1 & 0 & 0 & 0 & 0 & 0 & 0 & 0
        \\
        0 &  \pa_x & 0 & 1 & 0 & 0 & 0 & 0 & 0 & 0
        \\
        i m^1 {\tilde \psi}^{1}_{R} & 0 & 0 & 0 & 0 & 0 & 0 & 0 & 0 & 0
        \\
        i m^1 {\tilde \psi}^{1}_{L} & 0 & 0 & 0 & 0 & 0 & 0 & 0 & 0 & 0
        \\
        0 & i m^2 {\tilde \psi}^{2}_{R} & 0 & 0 & 0 & 0 & 0 & 0 & 0
& 0
        \\
        0 & i m^2 {\tilde \psi}^{2}_{L} & 0 & 0 & 0 & 0 & 0 & 0 & 0
& 0
        \end{array}\right), \nonumber
\er
\br
a_{12} = \left(\begin{array}{cccccccccc}
            \frac{im^{3}}{2} {\tilde \psi}^{3}_{R}
          & \frac{im^{3}}{2} {\tilde \psi}^{3}_{L} & 0 & 0 & m^1 {\psi}^{1}_{R}
                                                               & m^1 {\psi}^{1}_{L}
         & 0 & 0 & \frac{m^{3}}{2} {\psi}^{3}_{R} &
                   \frac{m^{3}}{2} {\psi}^{3}_{L}
        \\
        \frac{im^{3}}{2} {\tilde \psi}^{3}_{R} &
        \frac{im^{3}}{2} {\tilde \psi}^{3}_{L} & 0 & 0 & 0 & 0
        & m^2 {\psi}^{2}_{R}
        & m^2 {\psi}^{2}_{L}
        &\frac{m^{3}}{2} { \psi}^{3}_{R}
        &\frac{m^{3}}{2} { \psi}^{3}_{L}
        \\
        0 & 0 & 0 & 0 & 0 & 0 & 0 & 0 & 0 & 0
        \\
        0 & 0 & 0 & 0 & 0 & 0 & 0 & 0 & 0 & 0
        \\
        0 & 0 & -1 & 0 & 0 & 0 & 0 & 0 & 0 & 0
        \\
        0 & 0 & 0 & -1 & 0 & 0 & 0 & 0 & 0 & 0
        \\
        0 & 0 & 0 & 0 & -1 & 0 & 0 & 0 & 0 & 0
        \\
        0 & 0 & 0 & 0 & 0 & -1 & 0 & 0 & 0 & 0
        \\
        0 & 0 & 0 & 0 & 0 & 0 & -1 & 0 & 0 & 0
        \\
        0 & 0 & 0 & 0 & 0 & 0 & 0 & -1 & 0 & 0
        \end{array}\right), \nonumber
\er
\br
a_{21} &=& \left(\begin{array}{cccccccccc}
        \frac{im^{3}}{2} {\tilde \psi}^{3}_{R} &
        \frac{im^{3}}{2} {\tilde \psi}^{3}_{R} & 0 & 0 & 0 & 0 & 0 & 0 & 0& 0
        \\
        \frac{im^{3}}{2} {\tilde \psi}^{3}_{L} &
        \frac{im^{3}}{2} {\tilde \psi}^{3}_{L} & 0 & 0 & 0 & 0 & 0 & 0 & 0 & 0
        \\
        0 & 0 & 0 & 0 & 1 & 0 & 0 & 0 & 0 & 0
        \\
        0 & 0 & 0 & 0 & 0 & 1 & 0 & 0 & 0 & 0
        \\
        m^1 {\psi}^{1}_{R} & 0 & 0 & 0 & 0 & 0 & -1 & 0 & 0 & 0
        \\
        m^1 {\psi}^{1}_{L} & 0 & 0 & 0 & 0 & 0 & 0 & -1 & 0 & 0
        \\
        0 & m^2 {\psi}^{2}_{R} & 0 & 0 & 0 & 0 & 0 & 0 & -1 & 0
        \\
        0 & m^2 {\psi}^{2}_{L} & 0 & 0 & 0 & 0 & 0 & 0 & 0 & -1
        \\
        \frac{m^{3}}{2} {\psi}^{3}_{R} &
        \frac{m^{3}}{2} {\psi}^{3}_{R} & 0 & 0 & 0 & 0 & 0 & 0 & 0 & 0
        \\
        \frac{m^{3}}{2} {\psi}^{3}_{L} &
        \frac{m^{3}}{2} {\psi}^{3}_{L} & 0 & 0 & 0 & 0 & 0 & 0 & 0 & 0
        \end{array}\right),\nonumber
\\
a_{22}&=& \left(\begin{array}{ccccccc}
        0 & 0 & 0 & \cdots  & 0 & -1 & 0
        \\
        0 & 0 & 0 & \cdots  & 0 & 0 & -1
        \\
        0 & 0 & 0 & \cdots  & 0 & 0 & 0
        \\
        \multicolumn{7}{c}\dotfill
        \\
        0 & 0 & 0 &  \cdots  & 0 & 0
& 0
        \\
        -1 & 0 & 0 & \cdots  & 0 & 0
& 0
        \\
        0 & -1 & 0 & \cdots  & 0 & 0 & 0
        \end{array}\right). \nonumber
\er

The zero-modes are
\br
        {\bf {{v}}^{(2)}}^T(x) &=& \(u, \; \u, \; \o, \; \chi, \; 0, \; 0, \;
        m^{1} u\psi^{1}_{R}, \; m^{1} u\psi^{1}_{L}, \;
        m^{2} \u\psi^{2}_{R}, \; m^{2} \u\psi^{2}_{L}, \;
        \/{m^{3}}{2}(u+\u)\psi^{3}_{R},                                      \nn
        \\
        &&\/{m^{3}}{2}(u+\u)\psi^{3}_{L}, \; u^{\prime}+\o, \; \u^{\prime}+\chi, \;
        i m^{1}{\ti \psi}^{1}_{R}u, \; im^{1}{\ti \psi}^{1}_{L}u, \;
        im^{2} {\ti \psi}^{2}_{R}\u,                                                  \nn
        \\
        && im^{2} {\ti \psi}^{2}_{L}\u, \; \/{im^{3}}{2}{\ti \psi}^{3}_{R}(u+\u), \;
        i\/{m^{3}}{2}{\ti \psi}^{3}_{L}(u+\u)\).
\er

The zero-mode condition gives no constraints, implying the symmetries of the action
\br
\d {\bf \xi^{(2)}_{A}}\,=\, {\bf v^{(2)}_{A}}(x); \,\,\,\,A=1,2,...20.
\er

Now, let us choose the gauge conditions
\be
\lab{gauge2}
\O_{5}\equiv\pi_{1} J_{1}^{1}+\frac{1}{2}J_{1}.(j_{1}+ j_{3})=0, \qq  \O_{6}\equiv \pi_{2} J_{2}^{1}+\frac{1}{2} J_{2}.(j_{2}+ j_{3})=0,
\ee
and impose the consistency conditions with the Lagrange multipliers
$\eta^{5}, \eta^{6}$, then
\be
{\cal L}^{(3)} = \pi_{1}\dot{\phi}_{1} + \pi_{2}\dot{\phi}_{2} + \dot{\psi}_{R}\pi^{j}_{R}
        + \dot{\psi}_{L}\pi^{j}_{L}  + \dot{\eta}^{1}\O_{1} + \dot{\eta}^{2}\O_{2}
        + \dot{\eta}^{3}\O_{3} + \dot{\eta}^{4}\O_{4} + \dot{\eta}^{5}\O_{5}
        + \dot{\eta}^{6}\O_{6}- {\cal V}^{(3)},
\ee
where
\be
{\cal V}^{(3)}={\cal V}^{(2)}|_{\Omega_{5}=\Omega_{6}=0},
\ee
or explicitly
\br
{\cal V}^{(3)} &=&  \frac{1}{12}  \( (J_{1})^2 + (J_{2})^2 + J_{1}.J_{2} \)+\frac{1}{2} [ J_{1}.(j_{1}+ j_{3})+ J_{2}.(j_{2}+ j_{3}) ]
   \nn
        \\
        &+& i \tilde{\psi}^{j}_{R} {\psi}^{j}_{R,x}
         -  i \tilde{\psi}^{j}_{L} {\psi}^{j}_{L,x}
        + im^{j}_{\psi} \bar{\psi}^{j}\psi^{j}.
\er

The symplectic two-form for this Lagrangian is a non singular matrix, then our
algorithm has come to an end. Collecting the canonical part and the symplectic potential ${\cal V}^{(3)}$ one has
\be
\lab{thirring2}
{\cal L}[\psi,\overline{\psi}]= \sum_{j=1}^{3}\{i\overline{\psi}^{j}\gamma^{\mu}\pa_{\mu}\psi^{j}
        + m^{j}_{\psi}\,\,{\overline{\psi}}^{j}\psi^{j}\}\,
        \,-\sum_{\begin{array}{c} k,l=1\\k\leq l\end{array}}^{3}\[ \bar{a}_{kl} j_{k}.j_{l}\] + \sum_{l=1}^{3} m^{l}\nu_{l}\, j_{l}^{0},
\ee
where $\nu_{3} \equiv \frac{\nu_{1}+\nu_{2}}{2}$. We have made the same choice, $k=1/g$, and the field re-scalings $\psi^{j} \rightarrow 1/\sqrt{k}\,\psi^{j}$ as in the last section.
This is the same GMT Lagrangian as \rf{thirring1}. As a bonus, we get the chemical potentials\, $\mu_{l} \equiv m^{l} \nu_{l}$ ( $\dot{\eta}^{1, 2} \rightarrow \nu_{1,2 } $) times the charge densities.
These terms are related to the charges\, $Q_{F}^{l}=\frac{1}{2\pi}\int_{-\infty}^{+\infty} dx\, j_{l}^{0}(t,x)$, and
their presence is a consequence of the symplectic method \cite{annals}.

\subsection{The generalized sine-Gordon model (GSG)}
\label{sec:sG}

        One can choose other gauge fixings, instead of \rf{gauge1},
        to construct the twice-iterated Lagrangian. Let us make the choice
\be
\lab{gauge3}
\O_{3}\equiv J^{0}_{1} = 0, \qq  \O_{4}\equiv J^{0}_{2} = 0,
\ee
which satisfies the non-gauge invariance condition as can be verified by computing the
brackets $\{\O_{a} \; , \; J^{0}_{b}\}=0$;\, $a,b=1,2$. The twice-iterated Lagrangian
is obtained by bringing back these constraints into the canonical part of ${\cal L}^{(1)}$, then
\be
        {\cal L}^{(2)} = \pi_{1}\dot{\phi}_{1} + \pi_{2}\dot{\phi}_{2} + \dot{\psi}_{R}\pi^{j}_{R}
        + \dot{\psi}_{L}\pi^{j}_{L}  + \dot{\eta}^{1}\O_{1} + \dot{\eta}^{2}\O_{2}
        + \dot{\eta}^{3}\O_{3} + \dot{\eta}^{4}\O_{4} - {\cal V}^{(2)},
\ee
where the twice-iterated symplectic potential becomes
\be
{\cal V}^{(2)}={\cal V}^{(1)}\Big|_{\O_{3}= \O_{4} = 0}.
\ee

Considering the set of symplectic variables in the following order
\be
\xi^{(2)}_{A} = (\eta^{1}, \eta^{2},\eta^{3}, \eta^{4}, \phi_1, \phi_2, \psi^{1}_{R}, \psi^{1}_{L},
        \psi^{2}_{R}, \psi^{2}_{L}, \psi^{3}_{R}, \psi^{3}_{L}, \pi_{1}, \pi_2, \pi^{1}_{R},
        \pi^{1}_{L},
\pi^{2}_{R}, \pi^{2}_{L}, \pi^{3}_{R}, \pi^{3}_{L} )
\ee
and the components of the canonical one-form
\be
a^{(2)}_{A} = (\O_1, \O_2, \O_3, \O_4, \pi_{1}, \pi_2, -\pi^{1}_{R}, -\pi^{1}_{L}, -\pi^{2}_{R},
        -\pi^{2}_{L}, -\pi^{3}_{R}, -\pi^{3}_{L}, 0, 0, 0, 0, 0, 0, 0, 0),
\ee
the (degenerated) $20$x$20$ symplectic matrix is found to be
\br
f^{(2)}_{AB}(x,y) &=& \left(\begin{array}{cc}
        a_{11} &  a_{12}                        \\
        a_{21} &  a_{22}
        \end{array} \right) \d(x-y),
\er
where\\
$a_{11}=$
\br
\nn
 \left(\begin{array}{cccccccccc}
        0 & 0 &0 & 0 &  \pa_{x} & 0 & i m^1 {\tilde \psi}^{1}_{R}
        & i m^1 {\tilde \psi}^{1}_{L} & 0 & 0
        \\
        0 & 0 & 0 & 0 & 0 &  \pa_{x} & 0 & 0 & i m^2 {\tilde \psi}^{2}_{R}  &
        i m^2 {\tilde \psi}^{2}_{L}
        \\
        0 & 0 & 0 & 0 & 0 & 0 & i m^1 {\tilde \psi}^{1}_{R}
        & i m^1 {\tilde \psi}^{1}_{L} & 0 & 0
        \\
        0 & 0 & 0 & 0 & 0 & 0 & 0 & 0 & i m^2 {\tilde \psi}^{2}_{R}  &
        i m^2 {\tilde \psi}^{2}_{L}
        \\
         \pa_x & 0 & 0 & 0 & 0 & 0 & 0 & 0 & 0 & 0
        \\
        0 &  \pa_x & 0 & 0 & 0 & 0 & 0 & 0 & 0 & 0
        \\
        i m^1 {\tilde \psi}^{1}_{R} & 0 & i m^1 {\tilde \psi}^{1}_{R}
        & 0 & 0 & 0 & 0 & 0 & 0 & 0
        \\
        i m^1 {\tilde \psi}^{1}_{L} & 0 & i m^1 {\tilde \psi}^{1}_{L}
        & 0 & 0 & 0 & 0 & 0 & 0 & 0
        \\
        0 & i m^2 {\tilde \psi}^{2}_{R} & 0 & i m^2 {\tilde \psi}^{2}_{R}
        &  0 & 0 & 0 & 0 & 0 & 0
        \\
        0 & i m^2 {\tilde \psi}^{2}_{L} & 0 & i m^2 {\tilde \psi}^{2}_{L}
        & 0 & 0 & 0 & 0 & 0 & 0
        \end{array}\right), \nn
\er
\br
a_{12} = \left(\begin{array}{cccccccccc}
         \frac{im^{3}}{2} {\tilde \psi}^{3}_{R}
        & \frac{im^{3}}{2} {\tilde \psi}^{3}_{L} & 0 & 0 & m^1 {\psi}^{1}_{R}
        & m^1 {\psi}^{1}_{L} & 0 & 0 & \frac{m^{3}}{2} {\psi}^{3}_{R}
        & \frac{m^{3}}{2} {\psi}^{3}_{L}
        \\
        \frac{im^{3}}{2} {\tilde \psi}^{3}_{R}
        & \frac{im^{3}}{2} {\tilde \psi}^{3}_{L} & 0 & 0 & 0 & 0
        & m^2 {\psi}^{2}_{R} & m^2 {\psi}^{2}_{L}
        &\frac{m^{3}}{2} { \psi}^{3}_{R} &\frac{m^{3}}{2} { \psi}^{3}_{L}
        \\
        \frac{im^{3}}{2} {\tilde \psi}^{3}_{R}
        & \frac{im^{3}}{2} {\tilde \psi}^{3}_{L} & 0 & 0 & m^1 {\psi}^{1}_{R}
        & m^1 {\psi}^{1}_{L} & 0 & 0 & \frac{m^{3}}{2} {\psi}^{3}_{R}
        & \frac{m^{3}}{2} {\psi}^{3}_{L}
        \\
        \frac{im^{3}}{2} {\tilde \psi}^{3}_{R}
        & \frac{im^{3}}{2} {\tilde \psi}^{3}_{L} & 0 & 0 & 0 & 0
        & m^2 {\psi}^{2}_{R} & m^2 {\psi}^{2}_{L}
        &\frac{m^{3}}{2} { \psi}^{3}_{R} &\frac{m^{3}}{2} { \psi}^{3}_{L}
        \\
        0 & 0 & -1 & 0 & 0 & 0 & 0 & 0 & 0 & 0
        \\
        0 & 0 & 0 & -1 & 0 & 0 & 0 & 0 & 0 & 0
        \\
        0 & 0 & 0 & 0 & -1 & 0 & 0 & 0 & 0 & 0
        \\
        0 & 0 & 0 & 0 & 0 & -1 & 0 & 0 & 0 & 0
        \\
        0 & 0 & 0 & 0 & 0 & 0 & -1 & 0 & 0 & 0
        \\
        0 & 0 & 0 & 0 & 0 & 0 & 0 & -1 & 0 & 0
        \end{array}\right), \nn
\er
\br
a_{21} &=& \left(\begin{array}{cccccccccc}
        \frac{im^{3}}{2} {\tilde \psi}^{3}_{R} &
        \frac{im^{3}}{2} {\tilde \psi}^{3}_{R} &
        \frac{im^{3}}{2} {\tilde \psi}^{3}_{R} &
        \frac{im^{3}}{2} {\tilde \psi}^{3}_{R} & 0 & 0 & 0 & 0 & 0&0
        \\
        \frac{im^{3}}{2} {\tilde \psi}^{3}_{L} &
        \frac{im^{3}}{2} {\tilde \psi}^{3}_{L} &
        \frac{im^{3}}{2} {\tilde \psi}^{3}_{L} &
        \frac{im^{3}}{2} {\tilde \psi}^{3}_{L} & 0 & 0 & 0 & 0 & 0 &0
        \\
        0 & 0 & 0 & 0 & 1 & 0 & 0 & 0 & 0 &0
        \\
        0 & 0 & 0 & 0 & 0 & 1 & 0 & 0 & 0 &0
        \\
        m^1 {\psi}^{1}_{R} & 0 & m^1 {\psi}^{1}_{R} & 0 & 0 & 0 & -1 & 0 & 0 &0
        \\
        m^1 {\psi}^{1}_{L} & 0 & m^1 {\psi}^{1}_{L} & 0 & 0 & 0 & 0 &-1 & 0 &0
        \\
        0 & m^2 {\psi}^{2}_{R} & 0 & m^2 {\psi}^{2}_{R} & 0 & 0 & 0 & 0 & -1 &0
        \\
        0 & m^2 {\psi}^{2}_{L} & 0 & m^2 {\psi}^{2}_{L} & 0 & 0 & 0 & 0 & 0 &-1
        \\
        \frac{m^{3}}{2} {\psi}^{3}_{R} &
        \frac{m^{3}}{2} {\psi}^{3}_{R} &
        \frac{m^{3}}{2} {\psi}^{3}_{R} &
        \frac{m^{3}}{2} {\psi}^{3}_{R} & 0 & 0 & 0 & 0 & 0 &0
        \\
        \frac{m^{3}}{2} {\psi}^{3}_{L} &
        \frac{m^{3}}{2} {\psi}^{3}_{L} &
        \frac{m^{3}}{2} {\psi}^{3}_{L} &
        \frac{m^{3}}{2} {\psi}^{3}_{L} & 0 & 0 & 0 & 0 & 0 &0
        \end{array}\right), \nn\\
a_{22}& =& \left(\begin{array}{ccccccc}
        0 & 0 & 0 &\cdots & 0 & -1 & 0
        \\
        0 & 0 & 0 &\cdots & 0 & 0 & -1
        \\
        0 & 0 & 0 & \cdots & 0 & 0 & 0
        \\
        \multicolumn{7}{c}\dotfill
        \\
        0 & 0 & 0 &\cdots & 0 & 0 & 0
        \\
        -1 & 0 & 0 & \cdots & 0 & 0 & 0
        \\
        0 & -1 & 0 & \cdots & 0 & 0 & 0
        \end{array}\right).                    \nn
\er

Its zero-modes are
\br
\nn
{\bf {{v}}^{(2)}}^T(x) &=& \(u, \; \u, \; \o, \; \chi, \; 0, \; 0, \;
        m^{1}(u+\o)\psi^{1}_{R}, \;
        m^{1}(u+\o)\psi^{1}_{L}, \;\\\nn
&& m^{2}(\u+\chi)\psi^{2}_{R}, \;
        m^{2}(\u+\chi)\psi^{2}_{L},\,
       \/{m^{3}}{2}(u+\u+\o+\chi)\psi^{3}_{R}, \\\nn \;
  &&      \/{m^{3}}{2}(u+\u+\o+\chi)\psi^{3}_{L}, \;
        u^{\prime}, \; \u^{\prime}, \;
        im^{1}{\ti \psi}^{1}_{R}(u+\o),\\
                              \nn
        && im^{1}{\ti \psi}^{1}_{L}(u+\o), \;
        im^{2}{\ti \psi}^{2}_{R}(\u+\chi), \;
        im^{2}{\ti \psi}^{2}_{L}(\u+\chi), \;
        \\
&&
\/{im^{3}}{2}{\ti \psi}^{3}_{R}(u+\u+\o+\chi),
 \/{im^{3}}{2}{\ti \psi}^{3}_{L}(u+\u+\o+\chi)\),
\er
where $u$, $\u$, $\o$ and $\chi$ are arbitrary functions. The zero mode condition becomes
\br
\int {dx {{\bf v}^{(2)}}^{T}(x) \frac{\d}{\d \xi^{(2)}}\int{dy'{\cal V}^{(2)}}}
        = \int{dx\, J^{1}_{a}\, \partial_{x} f_{a}} \equiv 0,\,\,\,\,\,\,f_{a}\equiv (\o, \chi),\,\,\,a=1,2.\nn
\er

Since the functions $f_{a}$ are arbitrary we end up with the following constraints
\be
\lab{lagcons}
        \O_{5} \equiv J^{1}_1=0,\,\,\,\,\,\, \O_{6} \equiv J^{1}_2 = 0.
\ee

Notice that by solving the constraints, $\O_{3}=\O_{4}=\O_{5}=\O_{6}=0$, eqs. \rf{gauge3} and \rf{lagcons}, we may obtain
\be
\lab{majorana}
        {\ti \psi}^{j}_{R} = \psi^{j}_{R}, \qq {\ti \psi}^{j}_{L} = \psi^{j}_{L}.
\ee

So, at this stage, we have Majorana spinors, the scalars $\phi_{1}$ and $\phi_{2}$, and the auxiliary fields.
Next, introduce a third set of Lagrange multipliers into
${\cal L}^{(2)}$, then
\be
{\cal L}^{(3)} = \pi_{1}\dot{\phi}_{1} + \pi_{2}\dot{\phi}_{2} + \dot{\psi}_{R}\pi^{j}_{R}
        + \dot{\psi}_{L}\pi^{j}_{L}  + \dot{\eta}^{1}\O_{1} + \dot{\eta}^{2}\O_{2}
        + \dot{\eta}^{3}\O_{3} + \dot{\eta}^{4}\O_{4} + \dot{\eta}^{5}\O_{5}
        + \dot{\eta}^{6}\O_{6}- {\cal V}^{(3)},
\ee
where
\be
{\cal V}^{(3)}={\cal V}^{(2)}|_{\Omega_{5} = \Omega_{6} = 0}
\ee
or
\be
        {\cal V}^{(3)} = {1\over 24}{\phi_{j,x}}^2 + i{\psi}^{j}_{R}\psi^{j}_{R,x}
                    - i{\psi}^{j}_{L}\psi^{j}_{L,x}
                    + i m^{j}_{\psi}{\psi}^{j}_{R} \psi^{j}_{L} (e^{-i\phi_{j}}+ e^{i\phi_{j}}).
\ee

The new set of symplectic variables is assumed to be ordered as\\
$\xi^{(3)}_{A}=$
\br
 (\eta^{1}, \eta^{2},\eta^{3}, \eta^{4}, \eta^{5}, \eta^{6}, \phi_1,
        \phi_2, \psi^{1}_{R}, \psi^{1}_{L}, \psi^{2}_{R}, \psi^{2}_{L}, \psi^{3}_{R},
        \psi^{3}_{L}, \pi_{1}, \pi_2, \pi^{1}_{R}, \pi^{1}_{L},
\pi^{2}_{R}, \pi^{2}_{L},
        \pi^{3}_{R}, \pi^{3}_{L} ).\nn
\er

The components of the canonical one-form are\\
 $a_{A}^{(3)} =$
\br
 (\O_1, \O_2, \O_3, \O_4, \O_5, \O_6, \pi_{1}, \pi_2, -\pi^{1}_{R},
        -\pi^{1}_{L}, -\pi^{2}_{R}, -\pi^{2}_{L}, -\pi^{3}_{R}, -\pi^{3}_{L},
        0, 0, 0, 0, 0, 0, 0, 0).\nn
\er

After some algebraic manipulations we get the third-iterated $22$x$22$ symplectic two-form
\br
\lab{forma}
        f^{(3)}_{AB}(x,y) &=& \left(\begin{array}{cc}
        a_{11} &  a_{12}                        \\
        a_{21} &  a_{22}
        \end{array} \right) \d(x-y).
\er
where\\
$a_{11} =$
\br
{\tiny \left(\begin{array}{ccccccccccc}
        0 & 0 & 0 & 0 & 0 & 0 &  \pa_{x} & 0 & i m^1 {\tilde \psi}^{1}_{R}
        & i m^1 {\tilde \psi}^{1}_{L} & 0
        \\
        0 & 0 & 0 & 0 & 0 & 0 & 0 &  \pa_{x} & 0 & 0 & i m^2 {\tilde \psi}^{2}_{R}
        \\
        0 & 0 & 0 & 0 & 0 & 0 & 0 & 0 & i m^1 {\tilde \psi}^{1}_{R}
        & i m^1 {\tilde \psi}^{1}_{L} & 0
        \\
        0 & 0 & 0 & 0 & 0 & 0 & 0 & 0 & 0 & 0 & i m^2 {\tilde \psi}^{2}_{R}
        \\
        0 & 0 & 0 & 0 & 0 & 0 & 0 & 0 & -i m^1 {\tilde \psi}^{1}_{R}
        & i m^1 {\tilde \psi}^{1}_{L} & 0
        \\
        0 & 0 & 0 & 0 & 0 & 0 & 0 & 0 & 0 & 0 & -i m^2 {\tilde \psi}^{2}_{R}
        \\
         \pa_x & 0 & 0 & 0 & 0 & 0 & 0 & 0 & 0 & 0 & 0
        \\
        0 &  \pa_x & 0 & 0 & 0 & 0 & 0 & 0 & 0 & 0 & 0
        \\
        i m^1 {\tilde \psi}^{1}_{R} & 0 & i m^1 {\tilde \psi}^{1}_{R}
        & 0 & -i m^1 {\tilde \psi}^{1}_{R} & 0 & 0 & 0 & 0 & 0 & 0
        \\
        i m^1 {\tilde \psi}^{1}_{L} & 0 & i m^1 {\tilde \psi}^{1}_{L}
        & 0 & i m^1 {\tilde \psi}^{1}_{L} & 0 & 0 & 0 & 0 & 0 & 0
        \\
        0 & i m^2 {\tilde \psi}^{2}_{R} & 0 & i m^2 {\tilde \psi}^{2}_{R}
        & 0 & -i m^2 {\tilde \psi}^{2}_{R} & 0 & 0 & 0 & 0 & 0
        \end{array}\right)},                                                    \nn
\er\\
$a_{12}=$
\br
     {\tiny  \left(\begin{array}{ccccccccccc}
        0 & \frac{im^{3}}{2} {\tilde \psi}^{3}_{R}
        & \frac{im^{3}}{2} {\tilde \psi}^{3}_{L} & 0 & 0 & m^1 {\psi}^{1}_{R}
        & m^1 {\psi}^{1}_{L} & 0 & 0 & \frac{m^{3}}{2} {\psi}^{3}_{R}
        & \frac{m^{3}}{2} {\psi}^{3}_{L}
        \\
        i m^2 {\tilde \psi}^{2}_{L} & \frac{im^{3}}{2} {\tilde \psi}^{3}_{R}
        & \frac{im^{3}}{2} {\tilde \psi}^{3}_{L} & 0 & 0 & 0 & 0
        & m^2 {\psi}^{2}_{R} & m^2 {\psi}^{2}_{L}
        &\frac{m^{3}}{2} { \psi}^{3}_{R} &\frac{m^{3}}{2} { \psi}^{3}_{L}
        \\
        0 & \frac{im^{3}}{2} {\tilde \psi}^{3}_{R}
        & \frac{im^{3}}{2} {\tilde \psi}^{3}_{L} & 0 & 0 & m^1 {\psi}^{1}_{R}
        & m^1 {\psi}^{1}_{L} & 0 & 0 & \frac{m^{3}}{2} {\psi}^{3}_{R}
        & \frac{m^{3}}{2} {\psi}^{3}_{L}
        \\
        i m^2 {\tilde \psi}^{2}_{L} & \frac{im^{3}}{2} {\tilde \psi}^{3}_{R}
        & \frac{im^{3}}{2} {\tilde \psi}^{3}_{L} & 0 & 0 & 0 & 0
        & m^2 {\psi}^{2}_{R} & m^2 {\psi}^{2}_{L}
        &\frac{m^{3}}{2} { \psi}^{3}_{R} &\frac{m^{3}}{2} { \psi}^{3}_{L}
        \\
        0 & -\frac{im^{3}}{2} {\tilde \psi}^{3}_{R}
           & \frac{im^{3}}{2} {\tilde \psi}^{3}_{L} & 0 & 0
           & -m^1 {\psi}^{1}_{R} & m^1 {\psi}^{1}_{L} & 0 & 0
           & -\frac{m^{2}}{2} \psi^{3}_{R}
           &  \frac{m^{3}}{2} \psi^{3}_{L}
        \\
         i m^2 {\tilde \psi}^{2}_{L} & -\frac{im^{3}}{2} {\tilde \psi}^{3}_{R}
           & \frac{im^{3}}{2} {\tilde \psi}^{3}_{L} & 0 & 0 & 0 & 0
           & -m^2 {\psi}^{2}_{R} & m^2 {\psi}^{2}_{L}
           & -\frac{m^{3}}{2} \psi^{3}_{R}
           &  \frac{m^{3}}{2} \psi^{3}_{L}
        \\
        0 & 0 & 0 & -1 & 0 & 0 & 0 & 0 & 0 & 0 & 0
        \\
        0 & 0 & 0 & 0 & -1 & 0 & 0 & 0 & 0 & 0 & 0
        \\
        0 & 0 & 0 & 0 & 0 & -1 & 0 & 0 & 0 & 0 & 0
        \\
        0 & 0 & 0 & 0 & 0 & 0 & -1 & 0 & 0 & 0 & 0
        \\
        0 & 0 & 0 & 0 & 0 & 0 & 0 & -1 & 0 & 0 & 0
        \end{array}\right)},                            \nn
\er\\
$a_{21} =$
\br
{\tiny \left(\begin{array}{ccccccccccc}
        0 & i m^2{\tilde \psi}^{2}_{L} &
        0 & i m^2{\tilde \psi}^{2}_{L} & 0 & i m^2 {\tilde \psi}^{2}_{L} &
        0 & 0 & 0 & 0 & 0
        \\
        \frac{im^{3}}{2} {\tilde \psi}^{3}_{R} &
        \frac{im^{3}}{2} {\tilde \psi}^{3}_{R} &
        \frac{im^{3}}{2} {\tilde \psi}^{3}_{R} &
        \frac{im^{3}}{2} {\tilde \psi}^{3}_{R} &
       -\frac{im^{3}}{2} {\tilde \psi}^{3}_{R} &
       -\frac{im^{3}}{2} {\tilde \psi}^{3}_{R} &
        0 & 0 & 0 & 0 & 0
        \\
        \frac{im^{3}}{2} {\tilde \psi}^{3}_{L} &
        \frac{im^{3}}{2} {\tilde \psi}^{3}_{L} &
        \frac{im^{3}}{2} {\tilde \psi}^{3}_{L} &
        \frac{im^{3}}{2} {\tilde \psi}^{3}_{L} &
        \frac{im^{3}}{2} {\tilde \psi}^{3}_{L} &
        \frac{im^{3}}{2} {\tilde \psi}^{3}_{L} &
        0 & 0 & 0 & 0 & 0
        \\
        0 & 0 & 0 & 0 & 0 & 0 & 1 & 0 & 0 & 0 & 0
        \\
        0 & 0 & 0 & 0 & 0 & 0 & 0 & 1 & 0 & 0 & 0
        \\
        m^1 {\psi}^{1}_{R} & 0 &
        m^1 {\psi}^{1}_{R} & 0 &
       -m^1 {\psi}^{1}_{R} & 0 & 0 & 0 & -1 & 0 & 0
        \\
        m^1 {\psi}^{1}_{L} & 0 &
        m^1 {\psi}^{1}_{L} & 0 &
        m^1 {\psi}^{1}_{L} & 0 & 0 & 0 & 0 & -1 & 0
        \\
        0 & m^2 {\psi}^{2}_{R} & 0 &
            m^2 {\psi}^{2}_{R} & 0 &
            -m^2 {\psi}^{2}_{R} & 0 & 0 & 0 & 0 & -1
        \\
        0 & m^2 {\psi}^{2}_{L} & 0 &
            m^2 {\psi}^{2}_{L} & 0 &
           m^2 {\psi}^{2}_{L} & 0 & 0 & 0 & 0 & 0
        \\
        \frac{m^{3}}{2} {\psi}^{3}_{R} &
        \frac{m^{3}}{2} {\psi}^{3}_{R} &
        \frac{m^{3}}{2} {\psi}^{3}_{R} &
        \frac{m^{3}}{2} {\psi}^{3}_{R} &
       -\frac{m^{3}}{2} {\psi}^{3}_{R} &
       -\frac{m^{3}}{2} {\psi}^{3}_{R} & 0 & 0 & 0 & 0 & 0
        \\
        \frac{m^{3}}{2} {\psi}^{3}_{L} &
        \frac{m^{3}}{2} {\psi}^{3}_{L} &
        \frac{m^{3}}{2} {\psi}^{3}_{L} &
        \frac{m^{3}}{2} {\psi}^{3}_{L} &
        \frac{m^{3}}{2} {\psi}^{3}_{L} &
        \frac{m^{3}}{2} {\psi}^{3}_{L} & 0 & 0 & 0 & 0 & 0
        \end{array}\right)},                                            \nn
\er
\br
        a_{22} = \left(\begin{array}{ccccccccc}
        0 & 0 & 0 & 0 &\cdots & 0 & -1 & 0 & 0
        \\
        0 & 0 & 0 & 0 &\cdots & 0 & 0 & -1 & 0
        \\
        0 & 0 & 0 & 0 & \cdots & 0 & 0 & 0 & -1
        \\
        0 & 0 & 0 & 0 &\cdots & 0 & 0 & 0 & 0
        \\
        \multicolumn{9}{c}\dotfill
        \\
        0 & 0 & 0 & 0 &\cdots & 0 & 0 & 0 & 0
        \\
        -1 & 0 & 0 & 0 &\cdots & 0 & 0 & 0 & 0
        \\
        0 & -1 & 0 & 0 &\cdots & 0 & 0 & 0 & 0
        \\
        0 & 0 & -1 & 0 &\cdots & 0 & 0 & 0 & 0
        \end{array}\right).                    \nn
\er

It can be checked that this matrix has the zero-modes
\br
        {\bf {v}}^{(3)}(x) &=& (u, \; \u, \; \o, \; \chi, \; y, \; z, \; 0, \; 0, \;
        m^1 a^-_1\psi^1_R, \;
        m^1 a^+_1\psi^1_L, \;
        m^2 a^-_2\psi^2_R, \;
        m^2 a^+_2\psi^2_L,\nn \\
       \nn&& \frac{m^3}{2} a^{-}_3\psi^3_R,
         \frac{m^3}{2} a^{+}_3 \psi^3_L, \;  u^{\prime}, \;  \u^{\prime}, \;
        i m^1 a^-_1{\ti \psi}^1_R, \;
        i m^1 a^+_1{\ti \psi}^1_L, \;
        i m^2 a^-_2{\ti \psi}^2_R, \;
        i m^2 a^+_2{\ti \psi}^2_L, \;                      \nn
        \\
        && i\frac{m^3}{2} a^{-}_3{\ti \psi}^3_R, \;
        i\frac{m^3}{2} a^+_3{\ti \psi}^3_L ),
\er
where $a^+_1 \equiv u+\o+y,\, a^+_2 \equiv \u+\chi+z,\, a^+_3 \equiv  u+\o+y+\u+\chi+z,\, a^-_1 \equiv u+\o-y, \, a^-_2 \equiv \u+\chi-z, \, a^+_3 \equiv  u+\o+y+\u-\chi-z$, and $u$, $\u$, $\o$, $\chi$, $y$\, and $z$ are arbitrary functions. The relevant zero-mode condition gives no constraints. Then the action has the following symmetries
\br
\lab{symmetry1}
\d \xi^{(3)}_{A}\,=\, {\bf v^{(3)}_{A}}(x); \,\,\,\,A=1,2,...22.
\er

These symmetries allow us to fix the bilinears $i\psi^j_{R}\psi^j_{L}$ to be constants. By taking
$\psi_{R}^{j}=-iC_{j} \overline{\theta}_{j}$ and $\psi_{L}^{j}=\theta_{j}$, $(j=1,2,3)$ with $C_{j}$ being real numbers, we find that
$i\psi_{R}^{j}\psi_{L}^{j}$ indeed becomes a constant. Note that $\theta_{j}$ and $\overline{\theta}_{j}$ are
Grassmannian variables, while $\overline{\theta}_{j}\theta_{j}$ is an ordinary commuting number.

The two form $f_{AB}^{(3)}(x,y)$, eq. \rf{forma}, in the subspace
$(\phi_{1}, \phi_{2}, \pi_{\phi_{1}}, \pi_{\phi_{2}})$ defines a canonical symplectic structure modulo canonical transformations.
The coordinates $\phi_{a}$ and $\pi_{\phi_{a}}$\, ($a=1,2$) are not unique. Consider a canonical transformation from
$(\phi_{a}, \pi_{\phi_{a}})$ to $(\hat{\phi_{a}}, \hat{\pi}_{\hat{\phi_{a}}})$ such that
$\phi_{a}=\frac{\pa F}{\pa \pi_{\phi_{a}}}$ and $\hat{\phi_{a}}=\frac{\pa F}{\pa \hat{\pi}_{\hat{\phi}_{a}} }$. Then, in particular if $\phi_{a}=\hat{\phi_{a}}$ one can, in principle, solve for the
function $F$ such that a manifestly kinetic term appear in the new Lagrangian.

Then choosing $k=1/g$ as the overall coupling constant, we are left with
\be
\lab{sine}
{\cal L^{''}}=\sum_{j=1}^{3}\[\frac{1}{24 g} \partial_{\mu}\phi_{j} \partial^{\mu}\phi_{j}
        + \frac{M_{j}}{g} \; \mbox{cos} \phi_{j}\]+ \mu_{1}\pa_{x} \phi_{1}+\mu_{2} \pa_{x}\phi_{2},
\ee
where $M_{j}=m_{\psi}^{j}C_{j}$. This defines the {\sl  generalized
  sine-Gordon model} (GSG). In addition we get the terms multiplied by
chemical potentials $\mu_{1}$ and $\mu_{2}$ ($\dot{\eta}^{1,\,2} \rightarrow
-\mu_{1,\,2}$). These are just the topological charge densities, and are
related to the conservation of the number of kinks minus antikinks $Q_{\rm
  topol.}^{a}= \frac{1}{\pi} \int_{-\infty}^{+\infty} dx\,\,\pa_{x}
\phi_{a}$.

In the above gauge fixing procedures the possibility of Gribov-like ambiguities deserves a careful analysis. See ref. \cite{annals} for a discussion in the $sl(2)$ ATM case. However, in the next section, we provide an indirect evidence of the absence of such ambiguities, at least, for the soliton sector of the model.

Let us comment on the integrability of the GSG model. The usual SG model
 integrability can be shown by writing the model as a linear problem from
which the zero-curvature equation emerges as a compatibility condition. It
could be interesting to follow similar approach in the GSG case.

\section{The soliton/particle correspondences}
\label{sec:dual}

The $sl(2)$ ATM theory contains the sine-Gordon (SG) and the massive Thirring (MT) models
describing the soliton/particle correspondence of its spectrum \cite{bla,annals,witten}. The ATM one-(anti)soliton solution
satisfies the remarkable SG and MT classical correspondence in which, apart from the Noether and topological currents equivalence, MT spinor bilinears are related to the exponential of the SG field \cite{orfanidis}. The last relationship was exploited in \cite{nucl} to decouple the $sl(2)$ ATM equations of
motion into the SG and MT ones. Here we provide a generalization of that
correspondence to the $sl(3)$ ATM case. In fact, consider the relationships
\begin{eqnarray}
\nn
\frac{\psi _{R}^{1}\widetilde{\psi}_{L}^{1}}{i} &=&-\frac{1}{4\D}[\left(
m^{1}_{\psi} p_{1}-m^{3}_{\psi}p_{4}-m^{2}_{\psi}p_{5}\right) e^{i\left( \varphi _{2}-2\varphi_{1}\right) }+m^{2}_{\psi}p_{5}e^{3i\left( \varphi _{2}-\varphi _{1}\right) }  \\
&&+m^{3}_{\psi}p_{4}e^{-3i\varphi _{1}}-m^{1}_{\psi}p_{1}] \label{duality31}\\
\nn
\psi _{R}^{2}\widetilde{\psi}_{L}^{2} &=&-\frac{1}{4\D}[\left(
m^{2}_{\psi}p_{2}-m^{1}_{\psi}p_{5}-m^{3}_{\psi}p_{6}\right) e^{i\left( \varphi _{1}-2\varphi_{2}\right) }+m^{1}_{\psi}p_{5}e^{3i\left( \varphi _{1}-\varphi _{2}\right) }
\\
&&+m^{3}_{\psi}p_{6}e^{-3i\varphi _{2}}-m^{2}_{\psi}p_{2}]
\label{duality32} \\
\nn
\frac{\widetilde{\psi}_{R}^{3}\psi_{L}^{3}}{i} &=&-\frac{1}{4\D}[\left(
m^{3}_{\psi}p_{3}-m^{1}_{\psi}p_{4}+m^{2}_{\psi}p_{6}\right) e^{i\left( \varphi_{1}+\varphi_{2}\right) }+m^{1}_{\psi}p_{4}e^{3i\varphi _{1}}  \\
&&-m^{2}_{\psi}p_{6}e^{3i\varphi _{2}}-m^{3}_{\psi}p_{3}],  \label{duality33}
\end{eqnarray}
where $\D \equiv \bar{a}_{11}\bar{a}_{22}\bar{a}_{33}+2\bar{a}_{12}\bar{a}_{23}\bar{a}_{13}-\bar{a}_{11}\left(
\bar{a}_{23}\right)^{2}-\left( \bar{a}_{12}\right)^{2}\bar{a}_{33}-\left( \bar{a}_{13}\right)
^{2}\bar{a}_{22}$;\,  $p_{1}\equiv \left( \bar{a}_{23}\right)^{2}-\bar{a}_{22}\bar{a}_{33}$;\, $p_{2}\equiv \left( \bar{a}_{13}\right) ^{2}-\bar{a}_{11}\bar{a}_{33}$;\, $p_{3}\equiv \left(
\bar{a}_{12}\right)^{2}-\bar{a}_{11}\bar{a}_{22}$;\, $p_{4}\equiv \bar{a}_{12}\bar{a}_{23}-\bar{a}_{22}\bar{a}_{13}$;\, $p_{5}\equiv \bar{a}_{13}\bar{a}_{23}-\bar{a}_{12}\bar{a}_{33}$;\, $p_{6}\equiv \bar{a}_{11}\bar{a}_{23}-\bar{a}_{12}\bar{a}_{13}$ and the $\bar{a}_{ij}$'s being the current-current coupling constants of the GMT model \rf{thirring1}. The relationships \rf{duality31}-\rf{duality33} supplied with the conditions \rf{real1}-\rf{condi} and conveniently substituted into eqs. \rf{eqnm1}
and \rf{eqnm4}-\rf{eqnm15} decouple the $sl(3)^{(1)}$ CATM equations into the GSG \rf{sine} and GMT \rf{thirring1} equations of motion, respectively.

Moreover, one can show that the GSG \rf{sine} $M_{j}$ parameters and the GMT \rf{thirring1} couplings $\bar{a}_{ij}$ are related by
\begin{eqnarray}
\frac{2\D M_{1}}{g (m^{1}_{\psi})^{2}} &=&\bar{a}_{22}(-\frac{m^{3}_{\psi}}{m^{1}_{\psi}}
\bar{a}_{13}+\bar{a}_{33})+\bar{a}_{23}(-\bar{a}_{23}+\frac{m^{3}_{\psi}}{m^{1}_{\psi}}\bar{a}_{12}), \label{strongweak31} \\
\frac{2\D M_{2}}{g (m^{2}_{\psi})^{2}} &=&\bar{a}_{11}(-\frac{m^{3}_{\psi}}{m^{2}_{\psi}}
\bar{a}_{23}+\bar{a}_{33})+\bar{a}_{13}(-\bar{a}_{13}+\frac{m^{3}_{\psi}}{m^{2}_{\psi}}\bar{a}_{12}),  \label{strongweak32} \\
\frac{2\D M_{3}}{g (m^{3}_{\psi})^{2}} &=&-\frac{m^{1}_{\psi}m^{2}_{\psi}}{(m^{3}_{\psi})^{2}}
(\bar{a}_{12}\bar{a}_{33}-\bar{a}_{13}\bar{a}_{23})-\bar{a}_{11}\bar{a}_{22}+(\bar{a}_{12})^{2}. \label{strongweak33}
\end{eqnarray}

Various limiting cases of the relationships \rf{duality31}-\rf{duality33} and \rf{strongweak31}-\rf{strongweak33} are possible. First, let us consider
\br
\lab{limit1}
\bar{a}_{jk} \rightarrow \left\{\begin{array}{ll}\infty & \,\,\,j=k \neq l,\,\,\, (\mbox{for a given}\,\,\, l) \\
 \mbox{finite} & \,\,\,\mbox{other cases}
\end{array}
\right\}
\er
then one has
\br
\frac{\psi _{R}^{l}\widetilde{\psi}_{L}^{l}}{i}&=&\frac{m^{l}_{\psi}}{4}\left( e^{-i\phi_{l}}-1\right);\,\,\,\,\, \psi _{R}^{j}\widetilde{\psi}_{L}^{j}=0,\,\,\,\,\,\,\,\,j\neq l,
\lab{duality}
\er
for  $\bar{a}_{ll}=\d_{l} g \, (\d_{1,2}=1,\,\, \d_{3}=-1)$. The three species of one-soliton solutions of the $sl(3)$ ATM theory \rf{atm1}, found in \cite{bueno} and described in Section \ref{sec:ATM}, satisfy the relationships \rf{duality} \cite{nucl}. Moreover,
from eqs. \rf{strongweak31}-\rf{strongweak33} taking the same limits as
in \rf{limit1} one has
\begin{equation}
M_{l}=\frac{(m^{l}_{\psi})^{2}}{2}; \,\,\,\, M_{j}=0,\,\,\,j \neq l.  \lab{weakstrong}
\end{equation}

Therefore, the relationships \rf{duality31}-\rf{duality33} incorporate each $sl(2)$ ATM submodel (particle/soliton)
weak/strong coupling phases; i.e., the MT/SG correspondence \cite{nucl,annals}.

Then, the currents equivalence \rf{equivalence}, the relationships \rf{duality31}-\rf{duality33} and the conditions \rf{real1}-\rf{condi} satisfied by the one-soliton sector of CATM theory allowed us to establish the correspondence between the GSG and GMT models, thus extending the MT/SG result \cite{orfanidis}. It could be interesting to obtain the counterpart of eqs. \rf{duality31}-\rf{duality33} for the $N_{S} \geq 2$ solitons, for example along the lines of \cite{orfanidis}. For $N_{S}=2$, eq. \rf{equivalence} still holds \cite{bueno}; and eqs. \rf{real1}-\rf{condi} are satisfied for the species $(\a_{i},\a_{i})$.

Second, consider the limit
\br
\lab{limit2}
\bar{a}_{ik} \rightarrow \left\{\begin{array}{ll} \infty & \,\,\,i=k=j, \,\,\,\,(\mbox{for a chosen}\, j;\,\,\,j=1,2)\,\,\,
\\
 \mbox{finite} & \,\,\,\mbox{other cases}
\end{array}
\right\}
\er
one gets $M_{j}\,=\,0$ and
\br
4 \bar{\D}\frac{\psi _{R}^{l}\widetilde{\psi}_{L}^{l}}{i}&=& (m_{\psi}^{3}\bar{a}_{l3}-m_{\psi}^{l}\bar{a}_{33}) e^{-i\phi_{l}}-m_{\psi}^{3} \bar{a}_{l3} e^{-3i\varphi_{l}}+ m_{\psi}^{l} \bar{a}_{33},\,\,\, l\neq j \nn\\
4 \bar{\D}\frac{\psi _{R}^{3}\widetilde{\psi}_{L}^{3}}{i}&=&(m_{\psi}^{3}\bar{a}_{ll}-m_{\psi}^{l}\bar{a}_{l3})  e^{-i\phi_{3}}+m_{\psi}^{l} \bar{a}_{l3} e^{-3i\varphi_{l}}+ m_{\psi}^{3} \bar{a}_{ll},\\
\psi _{R}^{j}\widetilde{\psi}_{L}^{j}&=&0.
\lab{duality2}
\er
where $\bar{\D}\equiv 4(\bar{a}_{ll}\bar{a}_{33}-(\bar{a}_{l3})^2)$.  The parameters are related by $ (m_{\psi}^{3})^2 \bar{a}_{ll} M_{l} = m_{\psi}^{l}(m_{\psi}^{3}\bar{a}_{l3}-m_{\psi}^{l}\bar{a}_{33})  M_{3}$. In the case $M_{l}=M_{3}=M$ and redefining the fields as $\phi_{l}=\sqrt{12 g}(A+B),\, \phi_{j}=-\sqrt{12 g} B$ in the GSG sector, one gets the Lagrangian
\br
 {\cal L}_{BL}= \frac{1}{2}(\pa_{\mu} A)^2+ \frac{1}{2}(\pa_{\mu} B)^2+ 2\frac{M}{g} \mbox{cos}\sqrt{24 g} A\,\, \mbox{cos}\sqrt{72 g} B,
\er
which is a particular case of the Bukhvostov-Lipatov model (BL) \cite{bukhvostov}. It corresponds to a GMT-like theory with two Dirac spinors. The BL model is not classically integrable \cite{ameduri}, and some discussions have appeared in the literature about its quantum integrability \cite{saleur}.

Alternatively, if one allows the limit $\bar{a}_{33} \rightarrow \infty$ one gets $\psi _{R}^{3}\widetilde{\psi}_{L}^{3}\,=\,0$, and additional relations for the $\psi^{1}, \psi^{2}$ spinors and the $\varphi_{a}$ scalars. The parameters are related by $\frac{M_{1}}{(m_{\psi}^{1})^2 \bar{a}_{22}}=\frac{M_{2}}{(m_{\psi}^{2})^2 \bar{a}_{11}}=-\frac{M_{3}}{m_{\psi}^{1} m_{\psi}^{2} \bar{a}_{12}}$. Then we left with two Dirac spinors in the GMT sector and all the terms of the GSG model. The later resembles the $2-$cosine model studied in \cite{gerganov} in some sub-manifold of its renormalized parameter space.

\section{WZNW action and the GMT/GSG duality}

 In this section we re-derive the previous results in a full Lie
   algebraic  formulation. In section 4.3 an extension of the FJ method was
   applied to unravel the GSG sector. This method produced the loss of covariance which was restored by means of a complicated canonical transformation, so, here we overcome this problem applying the master Lagrangian method which also provides a final Lagrangian manifestly invariant under the relevant group symmetry of the scalar sector.  The off-critical $\hat{sl}(3)$ {\sl  affine Toda model coupled to
   matter fields} (ATM) is defined by the action \cite{jhep}
\br
\nn
\frac{1}{k} I_{\mbox{ATM}}^{(3)} &=& I_{WZNW}[b] + \int_{M}d^2x \{ \sum_{m=1}^{2} \Big[   <F_{m}^{-}\,,\, b F^{+}_{m} b^{-1}>\\
&&
\nn
-\frac{1}{2} <E_{-3}\, , \, [W^{+}_{m}\, , \, \pa_{+} W^{+}_{3-m}]> + <F_{m}^{-}\, , \, \pa_{+} W^{+}_{m}> \\
&&
+\frac{1}{2} <[W^{-}_{m}\,,\, \pa_{-} W^{-}_{3-m}]\,
 , \, E_{3}> + <\pa_{-} W^{-}_{m}
 \, , \, F_{m}^{+} > \Big]\},
\label{latm}
\er
where
\br
\label{wzw}
I_{WZNW}[b]\,=\, \frac{1}{8} \int_{M} d^2x Tr(\pa_{\mu}b \pa^{\mu}b^{-1}) + \frac{1}{12} \int_{D} d^3 x\, \epsilon^{ijk} Tr(b^{-1} \pa_{i}b b^{-1} \pa_{j}b b^{-1} \pa_{k}b),
\er
is the Wess-Zumino-Novikov-Witten (WZNW) action.

The $W$'s are some auxiliary fields and the $F$'s and  $E_{\pm 3}$ are given
in the Appendix. The field $b$ is given by
\br
\label{vp}
b\,=\,e^{i\vp_{1} H^{0}_{1}+i \vp_{2}H^{0}_{2}}
\er

From the action (\ref{latm}) we get the following matrix equations of motion:
\br
\label{auxiliar1}
\pa_{+} W_{m}^{+} &=& - b F^{+}_{m} b^{-1},\,\,\,\, \pa_{-} W_{m}^{-}\,=\, - b^{-1} F^{-}_{m} b,\\
\label{eqnw1}
\pa_{+}(\pa_{-} b b^{-1})&=&  \[E_{-3}\,,\, b E_{l}b^{-1}\] + \sum_{n=1}^{2} \[F_{n}^{-}\,,\, bF_{n}^{+}b^{-1}\]  \\
\label{eqnw2}
\pa_{-}F_{m}^{+}  &=& - \[E_{3}\,,\, \pa_{-} W^{-}_{3-m}\];\,\,\,\,
\pa_{+}F_{m}^{-} \,=\,  \[E_{-3}\,,\, \pa_{+} W^{+}_{3-m}\].
\er

Taking into account the eqs. (\ref{auxiliar1}) and the equations (\ref{eqnw2})
 we have the system of equations
\br
\label{eqq1}
\pa_{+}(\pa_{-} b b^{-1})& =&  \sum_{n=1}^{2} \[F_{n}^{-}\,,\, bF_{n}^{+}b^{-1}\]\\
\label{eqq2}
\pa_{-}F_{m}^{+}  &=&  \[E_{3}\,,\, b^{-1} F_{3-m}^{-} b\],\\
\label{eqq3}
\pa_{+}F_{m}^{-}  &=&-   \[E_{-3}\,,\, b F_{3-m}^{+} b^{-1}\],
\er
written in terms of the fields $b$ and $F^{\pm}_{m}$. The system of equations
(\ref{eqq1})-\ref{eqq3}) can be obtained from the zero-curvature equation (\ref{zeroc}) for (\ref{aa1})-(\ref{aa2})
with the replacement $B \rightarrow b$ provided that the fields
satisfy the constraints
\br
\label{sumconst}
  \[F_{2}^{+}\,,\, b^{-1}F_{1}^{-}b\]\,=\,0;\,\,\,  \[F_{2}^{-}\,,\, b F_{1}^{+} b^{-1}\]\,=\,0.
\er

The equations of motion  (\ref{eqnm1})-(\ref{eqnm16}) are
derived from a conformal version of the action (\ref{latm})
\cite{jhep}. However, by conveniently reducing the conformal model setting to
zero the field $\eta$, imposing the
reality conditions (\ref{real1}) or (\ref{real2}) and the constraints
(\ref{sumconst}), the subset of  equations (\ref{eqnm1}) and
(\ref{eqnm4})-(\ref{eqnm15}) can be obtained from the
system (\ref{eqq1})-(\ref{eqq3}).

It is an easy task to show
that the equations of motion (\ref{auxiliar1})-(\ref{eqnw2}) [ or
(\ref{eqq1})-(\ref{eqq3})], as well as the  constraints (\ref{sumconst}),
supplied with convenient transformations for the $ W_{m}^{\pm}$ fields, are
invariant under the left-right (\ref{leri1})-(\ref{leri2}) local symmetries.

\subsection{Generalized $\hat{sl}(3)$ sine-Gordon model}

In order to uncover the GSG sector of the model (\ref{latm}) we follow the
well known technique to perform duality mappings using the master action
approach \cite{hje}.  Following the approach one can remove the derivatives over $x_{\pm}$ in eqs. (\ref{eqnw2}) and write them as
\br
\label{eqnw21}
F_{m}^{+}  \,=\, - \[E_{3}\,,\, W^{-}_{3-m}\] - f_{m}^{+}(x_{+}),\,\,\,\,
F_{m}^{-}  \,=\,  \[E_{-3}\,,\,  W^{+}_{3-m}\] +  f_{m}^{-}(x_{-}),
\er
with $f_{m}^{\pm}(x_{\pm})$ being analytic functions. Making use of eqs. (\ref{eqnw21}) and (\ref{auxiliar1})  one can write the master action (\ref{latm}) as
\br
\nn
\frac{1}{k} I_{ATM}&=& I_{WZNW}(b) + \int_{M} \{-\frac{1}{2} \sum_{m=1}^{2} \Big[ 2<f_{m}^{-} b f_{m}^{+}  b^{-1}> +\\
&& <[E_{-3}\,,\,W^{+}_{3-m}] b f_{m}^{+}  b^{-1}>+ <[E_{3}\,,\,W^{-}_{3-m}] b^{-1} f_{m}^{-} b> \Big]\}
\label{ffs}
\er

Solving the auxiliary fields $f_{m}^{\pm}$ by means of their Euler-Lagrange equations of motion and replacing into the action (\ref{ffs}) one gets
\br
\frac{1}{k} I_{ATM}&=& I_{WZNW}(b) +\frac{1}{4} \int_{M} \{ \sum_{m=1}^{2} < [E_{3}\,,\, W_{3-m}^{-}] b^{-1} [E_{-3}\,,\, W_{3-m}^{+}]  b>\}.
\label{ffs1}
\er

In order to obtain the GSG sector we resort to a convenient gauge fixing
procedure. In fact, the action (\ref{ffs1}) is still invariant under the local
symmetries (\ref{leri1})-(\ref{leri2}). Therefore, one can make the
transformation:  $b \rightarrow b'$ and $[E_{\mp 3}\,,\,W^{\pm}_{3-m}]
\rightarrow \Lambda^{\mp}_{3-m}$, with $\Lambda^{\pm}_{3-m}$ being constant
elements in $\hat{{\cal G}}_{\pm (3-m)}$ with  similar structure to the $F^{\pm}_{3-m}$.
Then the gauge fixed version of (\ref{ffs1}) becomes
\br
\frac{1}{k} I_{ATM}&=& I_{WZNW}(b) + \int_{M} \[ \sum_{m=1}^{2}  <\Lambda_{m}^{-} b \Lambda_{m}^{+} b^{-1}>\].
\label{gsgm1}
\er
The equation of motion obtained from (\ref{gsgm1}) becomes
\br
\label{gsgm}
\pa_{+}(\pa_{-} b b^{-1})= \sum_{n=1}^{2} \[\L_{n}^{-}\,,\, b \L_{n}^{+}b^{-1}\].
\er

The eq. (\ref{gsgm}) defines the {\sl
  generalized sine-Gordon model (GSG)}. The parameters $\Lambda^{\pm}_{m}$
have dimension of mass and $k\,=\, \kappa/2\pi$,\, ($\kappa \in \IZ$).

Regarding the GMT sector, one can perform a symplectic method reduction
starting from the action (\ref{latm}) and taking into account the currents
equivalence (\ref{equivalence}) in matrix form. In dealing with the GMT sector
the FJ method turns out to be more appropriate, in  particular the Darboux's
transformation, which is inherent to this approach, simultaneously diagonalize
the spinor canonical one-forms, decouple these fields from the Toda fields,
and in addition maintains the algebraic structures of the spinor sector for
any affine Lie algebra (see ref. \cite{jhep} for more details). In the next
sub-section instead we will relate the
model (\ref{gsgm}) to a generalized massive Thirring model through a decoupling procedure.

\subsection{GMT/GSG mapping}

In this sub-section we compare the field equations of both dual models through
some mappings between the spinors (Dirac fields) and the  scalars (Toda
fields) using the Lie algebraic construction. Thus, we provide the MTM/GSG correspondence by decoupling the ATM eqs. of
motion (\ref{eqq1})-(\ref{eqq3}) into the eqs. of motion of the GMT
(\ref{thirring1}) [or (\ref{thirring2})] and GSG (\ref{sine}) models,
respectively.   In fact, consider the relationships \cite{jhep}
\br
\label{map1}
\sum_{n=1}^{2} \[F_{n}^{-}\,,\, bF_{n}^{+}b^{-1}\]  &= & \sum_{n=1}^{2}\[\L_{n}^{-}\,,\, b\L_{n}^{+}b^{-1}\] ,\\
\label{map2}
\[E_{-3}\,,\, bF_{3-m}^{+}b^{-1}\] &= &  \[E_{-3}\,,\,F_{3-m}^{+}\]+\frac{1}{16} \[\sum_{n} g_{mn} \[ F_{3-n}^{+}\,,\,W^{-}_{3-n} \]\,,\, F_{m}^{-} \],\\
\[E_{3}\,,\, b^{-1}F_{3-m}^{-} b \] &= &  \[E_{3}\,,\,F_{3-m}^{-}\] - \frac{1}{16} \[\sum_{n}  g_{mn} \[ F_{3-n}^{-}\,,\,W^{+}_{3-n} \]\,,\, F_{m}^{+} \],
\label{map3}
\\
\label{map4}
F^{\pm}_{m}&=& \mp  [E_{\pm 3}\,,\, W_{3-m}^{\mp}],\\
\label{sumconst1}
 \[F_{2}^{\pm}\,,\, b^{\mp}F_{1}^{\mp}b^{\pm}\]&=&0,
\er
where the eqs. (\ref{sumconst1}) are the constraints (\ref{sumconst}) written
in a compact form. The $\L_{n}^{\pm}$'s are some constant generators related
to the $M_{j}$ parameters in (\ref{sine}) and assumed to take
the same algebraic structure as the corresponding $F_{n}^{\pm}$. The $g_{mn}$
are some coupling constant parameters related to the $\bar{a}_{kl}$'s in (\ref{thirring1}).

The relationships (\ref{map2})-(\ref{map4}) when
conveniently substituted into (\ref{eqq2})-(\ref{eqq3}) provide
\br
\nn
\pa_{+}F_{m}^{-} \,=\,[E_{-3}\,,\,\pa_{+} W^{+}_{3-m}]&=& [E_{-3}\,,\,[E_{3}\,,W_{m}^{-}]]+\\
&&\frac{1}{16} [[E_{-3}\,,\, W^{+}_{3-m}]\,,\sum_{n=1}^{2} g_{mn} [[E_{3}\,,\, W^{-}_{n}]\,,\,  W^{-}_{3-n}\,]]\label{eqt1}
\\
\nn
-\pa_{-}F_{m}^{+} \,=\,[E_{3}\,,\,\pa_{-} W^{-}_{3-m}]&=&- [E_{3}\,,\,[E_{-3}\,,W_{m}^{+}]]-\\
&&\frac{1}{16} [[E_{3}\,,\, W^{-}_{3-m}]\,,\sum_{n=1}^{2} g_{mn}  [[E_{-3}\,,\, W^{+}_{n}]\,,\,  W^{+}_{3-n}\,]].
\label{eqt2}
\er

This system of equations is the matrix form of the GMT model (\ref{thirring1})
and can be derived from the action
\br
\nn
{\cal S}_{GMT}\,=\,\int d^2x  \sum_{m=1}^{2} \( \frac{1}{2} <[E_{-3}\, , \, W^{+}_{3-m}]\pa_{+} W^{+}_{m}> -
\frac{1}{2} <[ E_{3}\,,\,  W^{-}_{3-m}] \pa_{-} W^{-}_{m}]>-\\
 <[E_{-3}\,,\, W_{m}^{+}] [E_{3}\,,\, W_{m}^{-}] >\)-
 \frac{1}{2}  \sum_{m, n} < g_{mn} (j^{+})_{ m}\,  (j^{-})_{ n}>,
\label{gmt11}
\er
where the currents are given by
\br
\label{u1}
j^{0}_{m}&=&\frac{1}{8} \([[E_{-3}\,,\, W_{m}^{+}]\,,\, W_{3-m}^{+}] -
 [[E_{3}\,,\, W_{m}^{-}]\,,\, W_{3-m}^{-}] \)\\
\label{u11}
j^{1}_{m}&=&\frac{1}{8} \([[E_{-3}\,,\, W_{m}^{+}]\,,\, W_{3-m}^{+}] + [[E_{l}\,,\, W_{m}^{-}]\,,\, W_{3-m}^{-}]\).
\er

The coupling parameters $g_{mn}$ are symmetric by construction, then the current-current interaction terms can be put in a manifestly covariant form.
 The GMT model (\ref{gmt11}) includes the kinetic terms, the mass terms $<[E_{-3}\,,\, W_{m}^{+}] [E_{3}\,,\, W_{m}^{-}] >$, and the general covariant current-current interaction terms.  The canonical pairs are
\br
\nn
([E_{-3}\, , \, W^{+}_{3-m}] \,,\,  W^{+}_{m})\,\,\,\, \mbox{and}\,\,\,\, ([ E_{3}\,,\,  W^{-}_{3-m}] \,,\, W^{-}_{m }),
\er
where $ W^{\pm}_{m } \equiv$ Tr $\( W^{\pm}_{m}\, T^{(\mp m)}_{\l (m)}\)$\,;\, $ [E_{-3}\, , \, W^{+}_{3-m}] \equiv$ Tr $\( [E_{-3}\, , \, W^{+}_{3-m}]\, T^{(\pm m)}_{\l (m)}\)$.

Moreover, (\ref{map1}) when substituted into  (\ref{eqq1}) provides the GSG
system (\ref{gsgm}). The GSG system
(\ref{gsgm})-(\ref{gsgm1}) resemble the Leznov-Saveliev construction of the
usual SG theory. It could be an interesting problem to consider an analog
construction for the GSG model.

The Noether and topological currents equivalence must also be considered along
with the relationships (\ref{map1})- (\ref{sumconst1}), so the
eq. (\ref{equivalence}) can be written in the form
\br
\label{equivalence11}
\epsilon^{\mu \nu}\mbox{Tr} \( E^{0} b^{-1}\pa_{\nu} b \) \,=\, 4\mbox{Tr}\(E^{0} \sum_{m} j^{\mu}_{m} \)
\er
with  $E^{0}\equiv {\bf m}.H^{\pm}
=\frac{1}{6}[(2m^{1}_{\psi}+m^{2}_{\psi})H^{0}_{1}+(2m^{2}_{\psi}+m^{1}_{\psi})H^{0}_{2}]$
and the $U(1)$ currents $j^{\mu}_{m}$ in matrix form are given in (\ref{u1})-(\ref{u11})

The eq. (\ref{equivalence11}) must be considered as an additional relationship between the fields of the GMT and GSG theories.

Using (\ref{map2})-(\ref{map4}) and the constraints (\ref{sumconst1}) one can
establish a linear system of equations for some spinor bilinears with
``coefficients'' given by certain exponentials of the Toda fields, and by
solving this system of equations one obtains the spinor bilinears of the GMT
model (\ref{thirring1}) related to relevant $\vp_{a}$ fields of the GSG model
(\ref{sine}) which are presented in (\ref{duality31})-(\ref{duality33}). These relationships are analogous to those found in \cite{park} for fermion mass bilinears and bosonic ones for massive (free) non-Abelian fermions which are bosonized to certain symmetric space sine-Gordon models.
The spinor bilinears expressed in terms of exponentials of Toda fields
obtained in this way when conveniently substituted into  (\ref{map1}) will
provide us an identity equation for the scalars from which on can get some relationships between the parameters of both theories, the couplings $g_{mn}$, the fermion mass parameters of the GMT model
(\ref{eqt1})-(\ref{eqt1}) and  the $\L_{m}^{\pm}$ parameters of the GSG theory(\ref{gsgm}); i.e.
the $\bar{a}_{jk}$'s and the $M_{j}$ relationships as given in (\ref{strongweak31})-(\ref{strongweak33}).

\section{Generalization to higher rank Lie algebra}
\label{sec:sl(n)}

The procedures presented so far can directly be extended to the CATM model for the affine Lie algebra $sl(n)^{(1)}$ furnished with the principal gradation. According to the construction of \cite{matter}, these models have soliton solutions for an off-critical sub-model, possess a $U(1)$ vector current proportional to a topological current, apart from the conformal symmetry they exhibit a $\(U(1)_{R}\)^{n-1}\otimes \(U(1)_{L}\)^{n-1}$ left-right local gauge symmetry, and the equations of motion describe the dynamics of the scalar fields $\vp_{a},\, \eta,\, \widetilde{\nu}\, (a=1,...n-1)$ and the Dirac spinors $\psi^{\a_{j}}$, $\widetilde{\psi}^{\a_{j}}$, ($j=1,...N$; $N\equiv \frac{n}{2}(n-1)$ = number of positive roots\, $\a_{j}$\, of the simple Lie algebra $sl(n)$) with one-(anti)soliton solution associated to the field\, $\a_{j}.\vec{\vp}$\, ($\vec{\vp}=\sum_{a=1}^{n-1}\vp_{a} \a_{a}$,\, $\a_{a}$= simple roots of $sl(n)$)  for each pair of Dirac fields ($\psi^{\a_{j}}$, $\widetilde{\psi}^{\a_{j}}$)\cite{matter}. Therefore, it is possible to define the off-critical real Lagrangian $sl(n)$ ATM model for  the solitonic sector of the theory. The reality conditions would generalize the eqs. \rf{real1}-\rf{condi}; i.e., the new $\vp$'s real and the identifications $\widetilde{\psi}^{\a_{j}} \sim (\psi^{\a_{j}})^{*}$ (up to $\pm$ signs). To apply the symplectic analysis of $sl(n)$ ATM one must impose ($n-1$) constraints in the Lagrangian, analogous to \rf{lagrangian1}, due to the above local symmetries. The outcome will be a parent Lagrangian of a generalized massive Thirring model (GMT) with $N$ Dirac fields and a generalized sine-Gordon model (GSG) with ($n-1$) fields. The decoupling of the Toda fields and Dirac fields in the equations of motion of $sl(n)^{(1)}$ CATM, analogous to \rf{eqnm1}
and \rf{eqnm4}-\rf{eqnm15}, could be performed by an extension of the relationships \rf{duality31}-\rf{duality33} and \rf{real1}-\rf{condi}.

The Lagrangian formulation of the $sl(n)^{(1)}$ CATM model and its related
off-critical $\hat{sl}(n)$ ATM, as well as the weak-strong
duality formulation in matrix form have been presented in \cite{jhep}
thus providing the extension of the results in section 6.

\section{Outlook}
\label{sec:outlook}

In the context of the FJ, symplectic, and master Lagrangian approaches
we have shown that the $\hat{sl}(3)$ ATM \rf{atm1} theory is a parent
Lagrangian \cite{hje} from which both the GMT \rf{thirring1} and the GSG
\rf{sine} models are derivable. From \rf{thirring1} and \rf{sine}, it is also
clear the duality exchange of the couplings: $g \rightarrow 1/g$. The various
soliton/particle species correspondences were uncovered. The soliton sector satifies the $U(1)$ vector and topological currents equivalence \rf{equivalence} and decouples the equations of motion into both dual sectors, through the relationships \rf{duality31}-\rf{duality33} (supplied with \rf{real1}-\rf{condi}). The relationships \rf{duality31}-\rf{duality33} contain each $sl(2)$ ATM submodel soliton solution. In connection to these points, recently a parent Lagrangian method was used to give a generalization of the dual theories concept for non $p$-form fields \cite{casini}. In \cite{casini}, the parent Lagrangian contained both types of fields, from which each dual theory was obtained by eliminating the other fields through the equations of motion.

On the other hand, in nonabelian bosonization of massless fermions
\cite{witten1}, the fermion bilinears are identified with bosonic
operators. Whereas, in abelian bosonization \cite{coleman} there exists an
identification between the massive fermion operator or Mandelstam operator
(charge nonzero sector) and a non-perturbative bosonic soliton operator \cite{mandelstam}. Recently,
it has been shown that symmetric space sine-Gordon models bosonize the massive
nonabelian (free) fermions providing the relationships between the fermions
and the relevant solitons of the bosonic model \cite{park}. The ATM model
allow us to stablish these type of relationships for interacting massive
spinors in the spirit of particle/soliton correspondence. In this context, the
generalized Mandelstam soliton operators for the GMT model were recently constructed \cite{bosoni} and the fermion-boson mapping established through a set of
generalized bosonization rules in a quotient positive definite Hilbert space
of states. In addition, the
above approach to the GMT/GSG duality may be useful to construct the conserved
currents and the algebra of their associated charges in the context of the
CATM $\rightarrow$ ATM reduction. These currents in the MT/SG case were
constructed treating each model as a perturbation on a conformal field theory
(see \cite{kaul} and references therein).

The S-duality must be seen at the quantum level (the 2D analogue of the
Montonen-Olive conjecture). Regarding this issue the present work provides the
algebraic recipe to construct the bosonic dual of the GMT theory
(\ref{thirring1}) as has been recently performed in \cite{bosoni} by showing
that the
GSG model (\ref{sine}) structure remains almost the same in the full quantum regime.

Finally, using the explicit expressions for the effective actions (\ref{gsgm1})
and (\ref{gmt11}) corresponding to the GSG and GMT  models, respectively, one
can study the soliton properties, such as the time delays, in each dual
model. In addition, one can obtain the energy-momentum tensors, and hence compute in a direct way the masses of the solitons which were calculated in \cite{matter} as the result of the spontaneous breakdown of the conformal symmetry.

Moreover, two dimensional models with four-fermion interactions have played an important role in the understanding of QCD (see, e.g. \cite{bennett} and references therein). Besides, the GMT model contains explicit mass terms: most integrable models such as the Gross-Neveu, $SU(2)$ and $U(1)$ Thirring  models rather all present spontaneous mass generation, the exception being the massive Thirring model. A GMT submodel with $a_{ii}=0$, $a_{ij}=1\, (i>j)$ and equal $m_{\psi}^{j}$'s, defines the so-called extended Bukhvostov-Lipatov model (BL) and has recently been studied by means of a bosonization technique \cite{sakamoto1}. Finally, BL type models were applied to $N-$body problems in nuclear physics \cite{sakamoto}.

\vspace{1cm}

\noindent {\bf Acknowledgements}
\vspace{.3cm}

The author thanks the Mathematics Department of UFMT-ICET and
Prof. M.C. Ara\'ujo for hospitality and J. Acosta for
collaboration in a previous work. This work is supported by a
CNPq-FAPEMAT grant.

\appendix

\section{The $sl(3)^{(1)}$ CATM model }
\label{app:catm}
We summarize the construction and some properties of the CATM model relevant
to our discussions \footnote{Our notations follow that of \cite{jmp, jhep}.}. Consider the zero curvature condition
\br
\label{zeroc}
\partial_{+}A_{-}-\partial _{-}A_{+}+[A_{+},A_{-}]=0.
\er

The potentials take the form
\br
\label{aa1}
A_{+}=-BF^{+}B^{-1},\quad A_{-}=-\partial _{-}BB^{-1}+F^{-},\qquad
\er
with
\br
\label{aa2}
F^{+} \,=\,E^{3}+F_{1}^{+}+F_{2}^{+},\,\,\,\,\,\,
F^{-} \,=\, E^{-3}+F_{1}^{-}+F_{2}^{-},
\er
where $E^{\pm 3}\equiv {\bf m}.H^{\pm} =\frac{1}{6}[(2m^{1}_{\psi}+m^{2}_{\psi})H^{\pm 1}_{1}+(2m^{2}_{\psi}+m^{1}_{\psi})H^{\pm 1}_{2}]$ and the $F_{i}^{\pm }$'s and $B$ contain the spinor fields and scalars of the model, respectively
\br
\lab{F1}
F_{1}^{+}&=&\sqrt{im^{1}_{\psi}}\psi _{R}^{1}E_{\alpha _{1}}^{0}+\sqrt{im^{2}_{\psi}}\psi
_{R}^{2}E_{\alpha _{2}}^{0}+\sqrt{im^{3}_{\psi}}\widetilde{\psi }_{R}^{3}E_{-\alpha
_{3}}^{1},
\\
\lab{F2}
F_{2}^{+}&=&\sqrt{im^{3}_{\psi}}\psi _{R}^{3}E_{\alpha _{3}}^{0}+\sqrt{im^{1}_{\psi}}
\widetilde{\psi }_{R}^{1}E_{-\alpha _{1}}^{1}+\sqrt{im^{2}_{\psi}}\widetilde{\psi }
_{R}^{2}E_{-\alpha _{2}}^{1},
\\
\lab{F3}
F_{1}^{-}&=&\sqrt{im^{3}_{\psi}}\psi _{L}^{3}E_{\alpha _{3}}^{-1}-\sqrt{im^{1}_{\psi}}
\widetilde{\psi }_{L}^{1}E_{-\alpha _{1}}^{0}-\sqrt{im^{2}_{\psi}}\widetilde{\psi }
_{L}^{2}E_{-\alpha _{2}}^{0},
\\
\lab{F4}
F_{2}^{-}&=&\sqrt{im^{1}_{\psi}}\psi _{L}^{1}E_{\alpha _{1}}^{-1}+\sqrt{im^{2}_{\psi}}\psi
_{L}^{2}E_{\alpha _{2}}^{-1}-\sqrt{im^{3}_{\psi}}\widetilde{\psi }
_{L}^{3}E_{-\alpha _{3}}^{0},\\
B&=&e^{i\vp_{1} H^{0}_{1}+i\vp_{2} H^{0}_{2} }\,e^{\widetilde{\nu }C}\,e^{\eta
  Q_{ppal}}\equiv b\, e^{\widetilde{\nu }C}\,e^{\eta
  Q_{ppal}}, \lab{eq1}
\er
where $E_{\alpha _{i}}^{n},H^{n}_{1},H^{n}_{2}$ and  $C$ ($i=1,2,3; \, n=0,\pm 1$) are some generators of $sl(3)^{(1)}$; $Q_{ppal}$ being the principal gradation operator. The commutation relations for an affine Lie algebra in the Chevalley basis are
\br
&&\left[ \emph{H}_a^m,\emph{H}_b^n\right] =mC\frac{2}{\alpha_{a}^2}K_{a b}\delta _{m+n,0}  \lab{a7}\\
&&\left[ \emph{H}_a^m,E_{\pm \alpha}^n\right] = \pm K_{\alpha a}E_{\pm \alpha}^{m+n}
\lab{a8}\\
&&\left[ E_\alpha ^m,E_{-\alpha }^n\right]
=\sum_{a=1}^rl_a^\alpha \emph{H}_a^{m+n}+\frac 2{\alpha ^2}mC\delta
_{m+n,0}  \lab{a9}
\\
&&\left[ E_\alpha ^m,E_\beta ^n\right] =
\varepsilon (\alpha ,\beta )E_{\alpha +\beta }^{m+n};\qquad \mbox{if }\alpha
+\beta \mbox{ is a root \qquad }  \lab{a10}
\\
&&\left[ D,E_\alpha ^n\right] =nE_\alpha ^n,\qquad \left[ D,\emph{H}%
_a^n\right] =n\emph{H}_a^n.  \lab{a12}
\er
where $K_{\alpha a}=2\a.\a_{a}/\a_{a}^2=n_{b}^{\a}K_{ba}$, with $n_{a}^{\a}$ and $l_a^\alpha$ being the integers in the expansions $\a=n_{a}^{\a}\a_{a}$ and $\a/\a^2=l_a^\alpha\a_{a}/\a_{a}^2$, and $\varepsilon (\alpha ,\beta )$ the relevant structure constants.

Take $K_{11}=K_{22}=2$ and $K_{12}=K_{21}=-1$ as the Cartan matrix elements of the simple Lie algebra $sl(3)$. Denoting by $\a_{1}$ and $\a_{2}$ the simple roots and the highest one by $\psi (=\a_{1}+\a_{2})$, one has $l_{a}^{\psi}=1(a=1,2)$, and $K_{\psi 1}=K_{\psi 2}=1$. Take $\varepsilon (\alpha ,\beta )=-\varepsilon (-\alpha ,-\beta ),\,\, \varepsilon_{1,2}\equiv \varepsilon (\alpha_{1} ,\a_{2})=1,\,\,  \varepsilon_{-1,3}\equiv \varepsilon(-\alpha_{1} ,\psi )=1\,\, \mbox{and}\, \,\,\varepsilon_{-2,3}\equiv \varepsilon (-\alpha_{2} ,\psi)=-1$.

One has $Q_{ppal} \equiv \sum_{a=1}^{2}  {\bf s}_{a}\l^{v}_{a}.H + 3 D$, where $\l^{v}_{a}$ are the fundamental co-weights of $sl(3)$, and the principal gradation vector is ${\bf s}=(1,1,1)$ \cite{kac}.

The zero curvature condition (\ref{zeroc}) gives the following equations of motion
\br
\lab{eqnm1}
\frac{\partial ^{2}\varphi_{a}}{4i\,e^{\eta}} &=&m^{1}_{\psi}[e^{\eta -i\phi_{a}}\widetilde{\psi }_{R}^{l}\psi _{L}^{l}+e^{i\phi_{a}}\widetilde{\psi }_{L}^{l}\psi
_{R}^{l}]+m^{3}_{\psi}[e^{-i\phi_{3}}\widetilde{\psi }
_{R}^{3}\psi _{L}^{3}+e^{\eta +i\phi_{3}}\widetilde{\psi }
_{L}^{3}\psi _{R}^{3}];\,\,\,\,a=1,2\,\,\,\,\,\,\,\,\,\,\,\,\,\,\,\,\,\,\,
\\
\lab{eqnm3}
-\frac{\partial ^{2}\widetilde{\nu }}{4} &=&im^{1}_{\psi}e^{2\eta -\phi_{1}}\widetilde{\psi }_{R}^{1}\psi _{L}^{1}+im^{2}_{\psi}e^{2\eta
-\phi_{2}}\widetilde{\psi }_{R}^{2}\psi
_{L}^{2}+im^{3}_{\psi}e^{\eta -\phi_{3}}\widetilde{\psi }
_{R}^{3}\psi _{L}^{3}+{\bf m}^{2}e^{3\eta },\,\,
\\
\lab{eqnm4}
-2\partial _{+}\psi _{L}^{1}&=&m^{1}_{\psi}e^{\eta +i\phi_{1}}\psi
_{R}^{1},\,\,\,\,\,\,\,\,\,\,\,\,\,\,\,
-2\partial _{+}\psi _{L}^{2}\,=\,m^{2}_{\psi}e^{\eta +i\phi_{2}}\psi
_{R}^{2},
\\
\lab{eqnm5}
2\partial _{-}\psi _{R}^{1}&=&m^{1}_{\psi}e^{2\eta -i\phi_{1}}\psi
_{L}^{1}+2i \(\frac{m^{2}_{\psi}m^{3}_{\psi}}{im^{1}_{\psi}}\)^{1/2}e^{\eta }(-\psi _{R}^{3}
\widetilde{\psi }_{L}^{2}e^{i\phi _{2}}-
\widetilde{\psi }_{R}^{2}\psi _{L}^{3}e^{-i\phi_{3}}),
\\
\lab{eqnm7}
2\partial _{-}\psi _{R}^{2}&=&m^{2}_{\psi}e^{2\eta -i\phi_{2}}\psi
_{L}^{2}+2i\(\frac{m^{1}_{\psi}m^{3}_{\psi}}{im^{2}_{\psi}}\)^{1/2}e^{\eta }(\psi _{R}^{3}
\widetilde{\psi }_{L}^{1}e^{i\phi _{1}}+
\widetilde{\psi }_{R}^{1}\psi _{L}^{3}e^{-i\phi _{3}}),
\\
\lab{eqnm8}
-2\partial _{+}\psi _{L}^{3}&=&m^{3}_{\psi}e^{2\eta +i\phi _{3}}\psi
_{R}^{3}+2i\(\frac{m^{1}_{\psi}m^{2}_{\psi}}{im^{3}_{\psi}}\)^{1/2}e^{\eta }(-\psi _{L}^{1}\psi
_{R}^{2}e^{i\phi_{2}}+\psi _{L}^{2}\psi
_{R}^{1}e^{i\phi _{1}}),
\\
\lab{eqnm9}
2\partial _{-}\psi _{R}^{3}&=&m^{3}_{\psi}e^{\eta -i\phi_{3}}\psi
_{L}^{3},\,\,\,\,\,\,\,\,\,\,\,\,
2\partial _{-}\widetilde{\psi }_{R}^{1}\,=\,m^{1}_{\psi}e^{\eta +i\phi_{1}}\widetilde{\psi }_{L}^{1},
\\
\lab{eqnm10}
-2\partial _{+}\widetilde{\psi }_{L}^{1} &=&m^{1}_{\psi}e^{2\eta -i\phi_{1}}\widetilde{\psi }_{R}^{1}+2i\(\frac{m^{2}_{\psi}m^{3}_{\psi}}{im^{1}_{\psi}
}\)^{1/2}e^{\eta }(-\psi _{L}^{2}\widetilde{\psi }_{R}^{3}e^{-i\phi _{3}}-\widetilde{\psi }_{L}^{3}\psi _{R}^{2}e^{i\phi_{2}}),
\\
\lab{eqnm12}
-2\partial _{+}\widetilde{\psi }_{L}^{2} &=&m^{2}_{\psi}e^{2\eta -i\phi_{2}}\widetilde{\psi }_{R}^{2}+2i\(\frac{m^{1}_{\psi}m^{3}_{\psi}}{im^{2}_{\psi}}
\)^{1/2}e^{\eta }(\psi _{L}^{1}\widetilde{\psi }_{R}^{3}e^{-i\phi _{3}}+\widetilde{\psi }_{L}^{3}\psi _{R}^{1}e^{i\phi_{1}}),
\\
\lab{eqnm13}
2\partial _{-}\widetilde{\psi }_{R}^{2}&=&m^{2}_{\psi}e^{\eta+i\phi_{2}}\widetilde{\psi }_{L}^{2}, \,\,\,\,\,\,\,\,\,\,\,\,\,\,\,\,
-2\partial _{+}\widetilde{\psi }_{L}^{3}\,=\,m^{3}_{\psi}e^{\eta -i\phi _{3}}\widetilde{\psi }_{R}^{3},
\\
\lab{eqnm15}
2\partial _{-}\widetilde{\psi }_{R}^{3} &=&m^{3}_{\psi}e^{2\eta +i\phi_{3}}\widetilde{\psi }_{L}^{3}+2i\(\frac{m^{1}_{\psi}m^{2}_{\psi}}{im^{3}_{\psi}}
\)^{1/2}e^{\eta }(\widetilde{\psi} _{R}^{1}\widetilde{\psi }_{L}^{2}e^{i\phi_{2}}-\widetilde{\psi }_{R}^{2}\widetilde{\psi }_{L}^{1}e^{i\phi_{1}}),
\\
\lab{eqnm16}
\partial^{2}\eta&=&0,
\er
where $\phi_{1}\equiv2 \vp_{1}-\vp_{2},\,\phi_{2}\equiv 2\vp_{2}-\vp_{1},\,\phi_{3} \equiv \vp_{1}+\vp_{2}$.

Apart from the {\sl conformal invariance} the above equations exhibit the $\(U(1)_{L}\)^{2}\otimes \(U(1)_{R}\)^{2}$ {\sl left-right local gauge symmetry}
\br
\lab{leri1}
\vp_{a} &\ra& \vp_{a} + \theta_{+}^{a}( x_{+}) + \theta_{-}^{a}( x_{-}),\,\,\,\,a=1,2\\
\widetilde{\nu} &\ra& \widetilde{\nu}\; ; \qquad \eta \ra \eta \\
\psi^{i} &\ra & e^{i( 1+ \gamma_5) \Theta_{+}^{i}( x_{+})
+ i( 1- \gamma_5) \Theta_{-}^{i}( x_{-})}\, \psi^{i},\\
\,\,\,\,
\widetilde{\psi}^{i} &\ra& e^{-i( 1+ \gamma_5) (\Theta_{+}^{i})( x_{+})-i ( 1- \gamma_5) (\Theta_{-}^{i})( x_{-})}\,\widetilde{\psi}^{i},\,\,\, i=1,2,3;\lab{leri2}
\\
&&\Theta^{1}_{\pm}\equiv \pm \theta_{\pm}^{2} \mp 2\theta_{\pm}^{1},\,\,\Theta^{2}_{\pm}\equiv \pm \theta_{\pm}^{1}\mp 2\theta_{\pm}^{2},\,\,\Theta_{\pm}^{3}\equiv \Theta_{\pm}^{1}+\Theta_{\pm}^{2}.\nn
\er

One can get global symmetries for $\theta_{\pm}^{a}=\mp \theta_{\mp}^{a}=$
constants. For a model defined by a Lagrangian these would imply the presence
of two vector and two chiral conserved currents. However, it was found only
half of such currents \cite{bueno}. This is a consequence of the lack of a
Lagrangian description for the $sl(3)^{(1)}$ CATM in terms of the $B$ and
$F^{\pm}$ fields. However, for the sub-model Lagrangian (\ref{latm}) it is possible to construct the Noether currents for
each global symmetry.

The gauge fixing of the conformal symmetry, by setting the field $\eta$ to a constant, is used to stablish the $U(1)$ vector, $J^{\mu}=\sum_{j=1}^{3} m^{j}_{\psi} \bar{\psi}^{j}\gamma^{\mu}\psi^{j}$, and topological currents equivalence \cite{matter,annals}. Moreover, it has been shown that the soliton solutions are in the orbit of the solution $\eta=0$. The remarkable equivalence is
\br
\lab{equivalence}
\sum_{j=1}^{3} m^{j}_{\psi} \bar{\psi}^{j}\gamma^{\mu}\psi^{j} \equiv \epsilon^{\mu \nu}\partial_{\nu} (m^{1}_{\psi}\varphi_{1}+m^{2}_{\psi}\varphi_{2}),\,\,\,\,\,\,\, m^{3}_{\psi}=m^{1}_{\psi}+ m^{2}_{\psi},\,\,\,\,m^{i}_{\psi}>0.
\er

\end{document}